\documentclass[twocolumn,noshowpacs,pre,preprintnumbers,superscriptaddress,amsmath,amssymb,floatfix]{revtex4}
\usepackage[caption=false]{subfig}
\usepackage{graphicx}
\usepackage{color}
\usepackage{placeins}
\newcommand{\kk}{\langle k \rangle}
\newcommand{\er}{Erd\H{o}s-R\'{e}nyi }

\begin{document}
\title{Two distinct transitions in  spatially embedded multiplex networks}
\author{Michael M. Danziger}
\affiliation{Department of Physics, Bar Ilan University, Ramat Gan, Israel}
\author{Louis M. Shekhtman}
\affiliation{Department of Physics, Bar Ilan University, Ramat Gan, Israel}
\author{Yehiel Berezin}
\affiliation{Department of Physics, Bar Ilan University, Ramat Gan, Israel}
\author{Shlomo Havlin}
\affiliation{Department of Physics, Bar Ilan University, Ramat Gan, Israel}
\date{\today}
\begin{abstract}
Multilayer infrastructure is often interdependent, with nodes in one layer depending on nearby nodes in another layer to function.
The links in each layer are often of limited length, due to the construction cost of longer links.
Here, we model such systems as a multiplex network composed of two or more layers, each with links of characteristic geographic length, embedded in 2-dimensional space.
This is equivalent to a system of interdependent spatially embedded networks in two dimensions in which the connectivity links are constrained in length but varied while the length of the dependency links is always zero.
We find two distinct percolation transition behaviors depending on the characteristic length, $\zeta$, of the links. 
When $\zeta$ is longer than a certain critical value, $\zeta_c$, abrupt, first-order transitions take place, while for $\zeta<\zeta_c$ the transition is continuous.
We show that, though in single-layer networks increasing $\zeta$ decreases the percolation threshold $p_c$,  
in multiplex networks it has the opposite effect: increasing  $p_c$ to a maximum at $\zeta=\zeta_c$. 
By providing a more realistic topological model for spatially embedded interdependent and multiplex networks and highlighting its similarities to lattice-based models, we provide a new direction for more detailed future studies.
\end{abstract}
\maketitle

\section{Introduction}
Several models have been proposed for spatially embedded networks \cite{doar1993bad, wei1993comparison,  zegura1997quantitative, watts-nature1998,penrose2003random, kleinberg2000small,kosmidis-epl2008,li-epl2011, barthelemy-physicsreports2011,mcandrew2015robustness}.
In lattice-based models,  links are only formed to nearest or next nearest neighbors.
In random geometric models,  links are formed to all neighbors within some distance \cite{rozenfeld-prl2002, bradonjic2007giant}.
In models of power grid topology,  links are formed with the $m$ nearest neighbors, statically \cite{hines2010topological} or as a generative model \cite{deka-sitis2013}.
Some models utilize a cost function \cite{manna-jphysicsa2003,gastner2006spatial, emmerich2014structural,ren-naturecomm2014} or a characteristic distance distribution \cite{wang-proceedings2008,grassberger2013sir} to determine link lengths.
The model which we study here has spatiality expressed via characteristic link lengths.
We utilize exponentially distributed link lengths, similar to the Waxman model \cite{waxman1988routing}.

\begin{figure}
\includegraphics[width=0.93\linewidth]{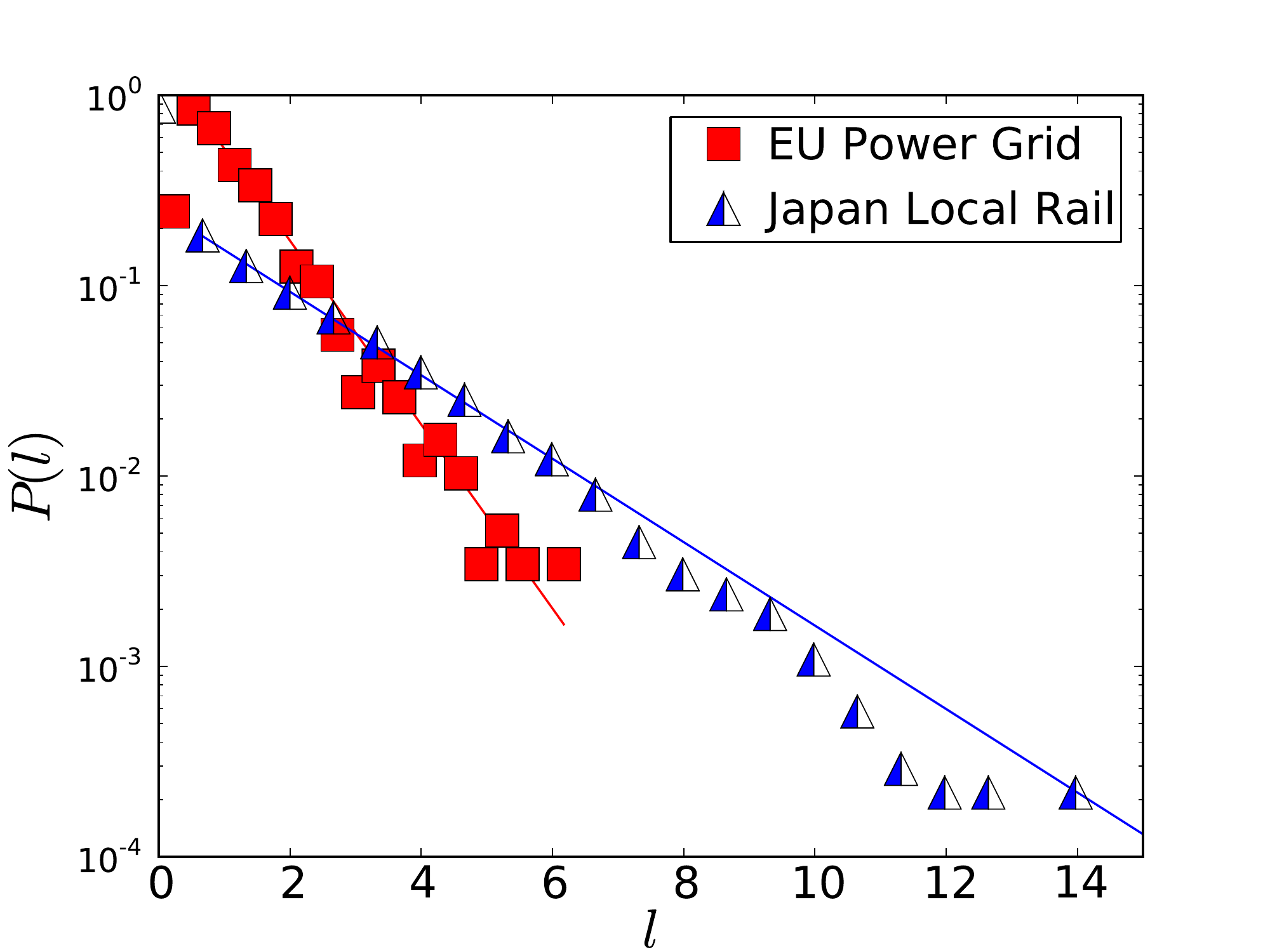}
\caption{\textbf{Examples of real-world networks with links of characteristic length.} 
We examine the distribution of the geographic lengths of the edges in both the European power grid \cite{zhou-ieee2005} (1851 edges) and the inter-station local railway lines in Japan \cite{japanrail} (20745 edges). 
These networks have links of characteristic length and longer links are exponentially unlikely, as indicated by the linear drop on the semi-logarithmic plot.
To compare the two datasets, we rescale the lengths so that they are measured in units of their own minimum length, which we determine as the peak of the distribution (mode length).
The normalization value ($l=1$) corresponds to 3.7 km (power) and 1.0 km (rail).
The characteristic length as determined by the mean is 4.8 km (power) and 1.2 km (rail).
As the slope of the exponential fit, it is 3.3 km (power) and 2.0 km (rail).
The Japan local railway data is formed from the complete railway network from \cite{japanrail} with bullet train lines and internal station tracks removed.}
\label{fig:realworld}
\end{figure}

\begin{figure*}
\subfloat[]{\includegraphics[width=0.4\linewidth]{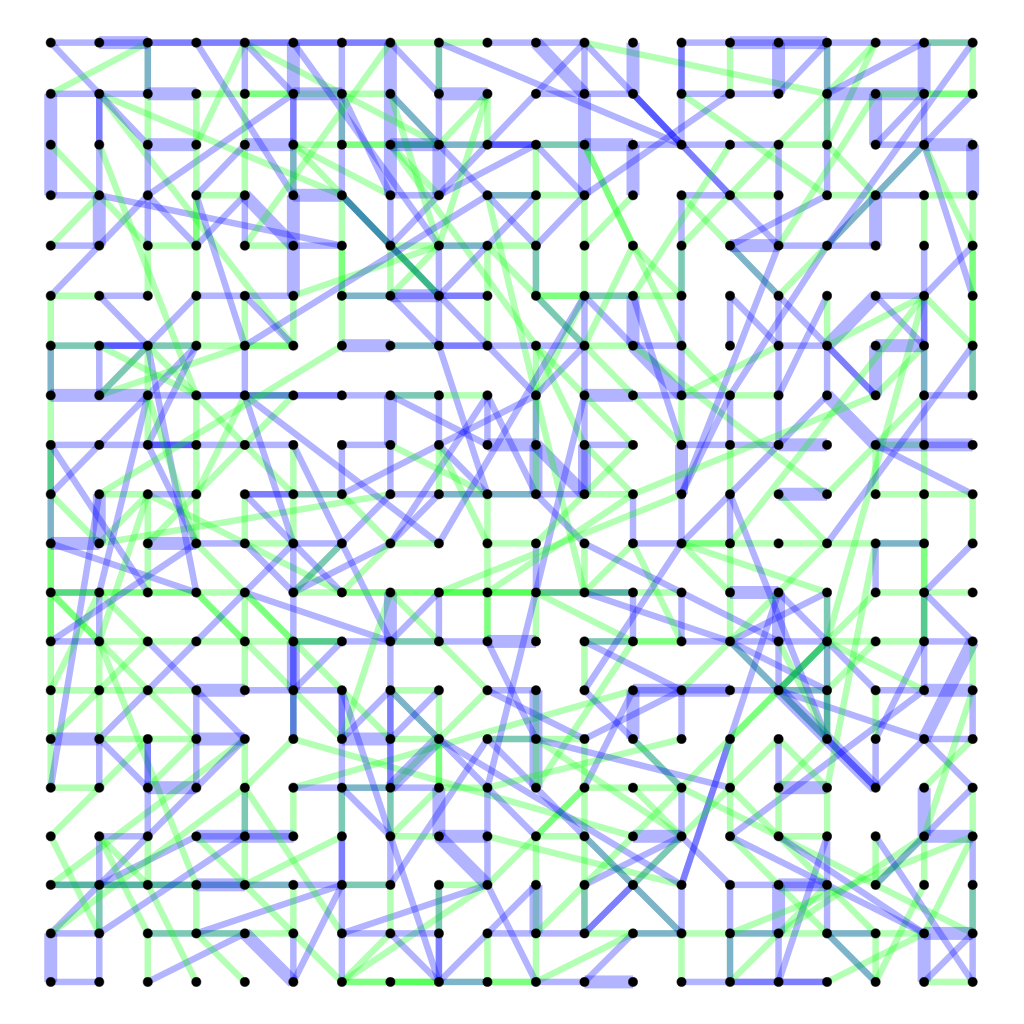} 
\label{fig:layers}}
\subfloat[]{\includegraphics[width=0.4\linewidth]{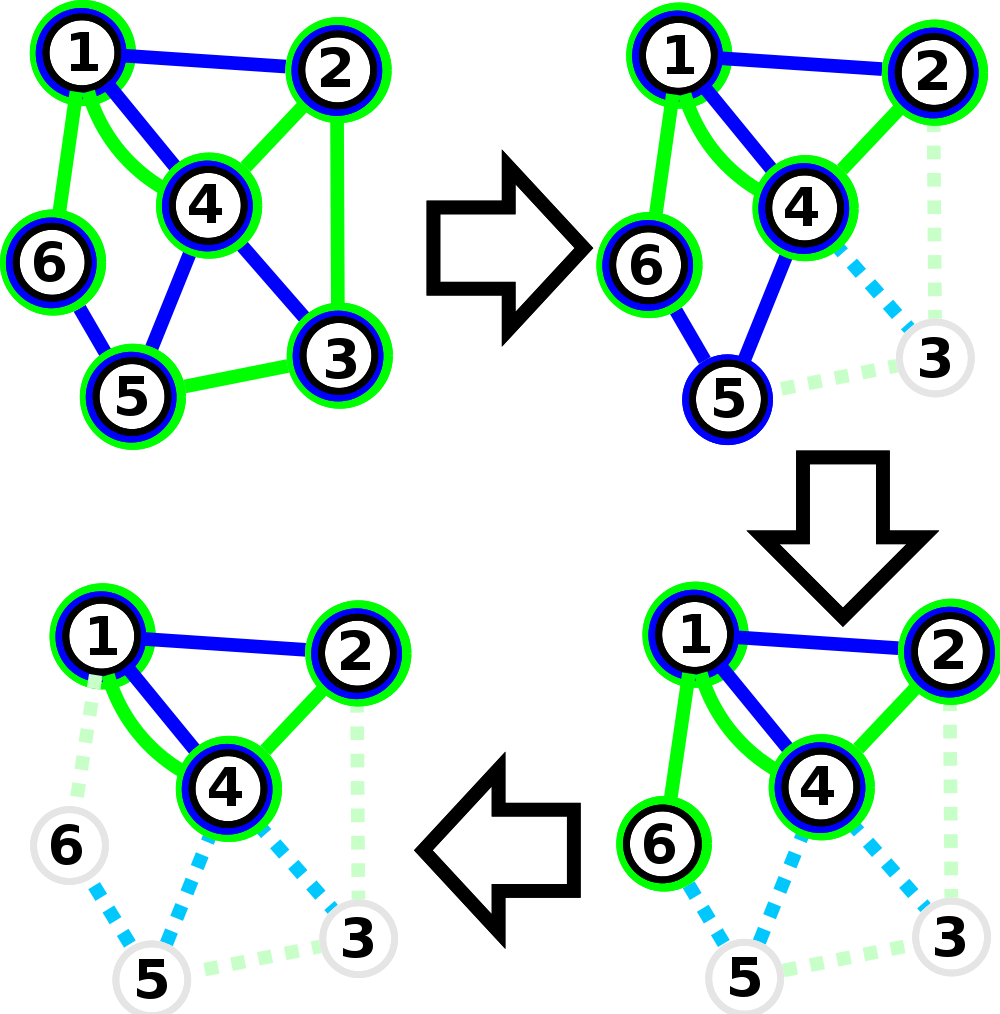}
\label{fig:cascade-diagram}}
\caption{\textbf{Spatially embedded multiplex networks.} 
\textbf{(a)}  The nodes occupy regular locations in two-dimensional space while the links in each layer (blue and green) have lengths that are exponentially distributed with characteristic length $\zeta=3$ and are connected at random. 
\textbf{(b) Cascading failures in multiplex networks.} 
In the first stage the mutual giant connected component (MGCC) consists of the entire network since all of the nodes are in the giant component of both layers (blue links and green links). An initial attack on Node 3 causes it and its links to fail. 
This detaches Node 5 from the giant component of the green links, and in the next step it and its links fail. 
After the failure of Node 5, Node 6 is no longer in the giant component of the blue links. 
After Node 6's failure we find that the remaining nodes are in the giant component of both layers. 
We note that the MGCC is not simply the intersection of the giant components in the separate layers. 
For example, using such an approach one would conclude that Node 6 is in the MGCC  after the failure of Node 3, which is not the case. 
}
\end{figure*}
Interdependent networks have been studied mainly on random topologies where analytic calculations are possible \cite{buldyrev-nature2010,gao-naturephysics2012,peixoto-prl2012,radicchi-naturephysics2013,kivela-jcomnets2014,boccaletti-physicsreports2014}.
However, since many real-world complex systems are embedded in space, it is important to understand the properties of interdependent networks with topology reflecting the dimensionality of the space \cite{gastner2006spatial,li-naturephysics2011}.
This is particularly important when dealing with critical infrastructure which is heavily influenced by spatial constraints \cite{rinaldi-ieee2001,barthelemy-physicsreports2011,hokstad-book2012,helbing-nature2013}.
An important first step in that direction is the model presented by Li et al. \cite{wei-prl2012} which models the networks as lattices and includes dependency links which are of limited geographic length (described by the parameter $r$).
This model was shown to include a number of surprising properties including three regimes of phase transitions depending on the value of $r$.  
For $r<r_c$, the phase transition is second-order and appears to be in the same universality class as 2-dimensional lattices \cite{danziger-newjphysics2015}.  
The percolation threshold increases as $r$ increases and reaches a maximum value at $r_c$, where the transition switches to first-order \cite{wei-prl2012,danziger-jcomnets2014}.
For $r_c<r<\infty$ the transition is first-order and characterized by a spreading process, with $p_c$ decreasing monotonically. 
For $r=\infty$, the transition is an abrupt simultaneous first and second order transition \cite{danziger-jcomnets2014,dong-pre2014}.
The model was studied under partial dependency with $r=\infty$ and it was shown to have first-order transitions for any fraction of dependency \cite{bashan-naturephysics2013}.
When there is partial dependency and finite dependency link lengths, $r_c$ is shown to increase as the fraction of dependent nodes decreases, diverging at $q=0$ \cite{danziger-sitis2013,danziger-jcomnets2014}.
This model was also studied for general networks formed of interdependent lattices \cite{shekhtman-pre2014} for interdependent resistor networks with process-based dependency \cite{danziger-newjphysics2015} and in the presence of healing \cite{stippinger-physa2014}. 

However, in many real-world systems, the length of the dependency links may not be longer than the length of the connectivity links.
For instance, in the example of the power grid and communications network, it is unlikely that a communications station will skip the nearer power stations and depend on a power station that is farther away. 
Also,  real-world networks including the power grid do not have uniform link lengths like a lattice, but rather have links of a characteristic length, consistent with an exponential distribution (see example in Fig. \ref{fig:realworld}) \cite{li-naturephysics2011}.
To address these two issues, here we model interdependent networks where every node has a bi-directional dependency link with the nearest geographic node in the other network.
We treat each pair of nodes as single nodes in a multiplex network, where the links in each layer are different but of the same characteristic length $\zeta$.
We thus interpret each node as a geographic entity (e.g. a city or neighborhood) which is linked via two types of links (e.g. electricity and communications) to other nodes.
Each node requires both of its constituents to function and each constituent requires connectivity within its layer.

Our main focus in this paper is to examine the role of the characteristic length ($\zeta$) of the links on the robustness of the multiplex.
We find that, though increasing $\zeta$  \textit{decreases} $p_c$ in single networks--making them more robust--it has the opposite effect on multiplex networks.
Increasing $\zeta$ \textit{increases} $p_c$ for multiplex networks until a critical length $\zeta_c$ where $p_c$ is maximal.
At $\zeta_c$, the percolation transition changes to first-order and  the multiplex network is susceptible to spreading cascading failures similar to the ones observed in interdependent lattices \cite{wei-prl2012,danziger-sitis2013,danziger-jcomnets2014,berezin-scireports2015}.

By demonstrating that comparable critical behavior emerges in more realistic topologies, we show that the critical behavior demonstrated in previous lattice-based models is not limited to the specific implementation or the lattice topology but is rather a generic property of interdependent spatially embedded networks.
Furthermore, by providing a topological model that more closely matches real-world systems, we provide a more realistic topological framework for future studies of interdependent critical infrastructure.

\begin{figure*}
\centering
\subfloat[single-layer]{\includegraphics[width=0.4\textwidth]{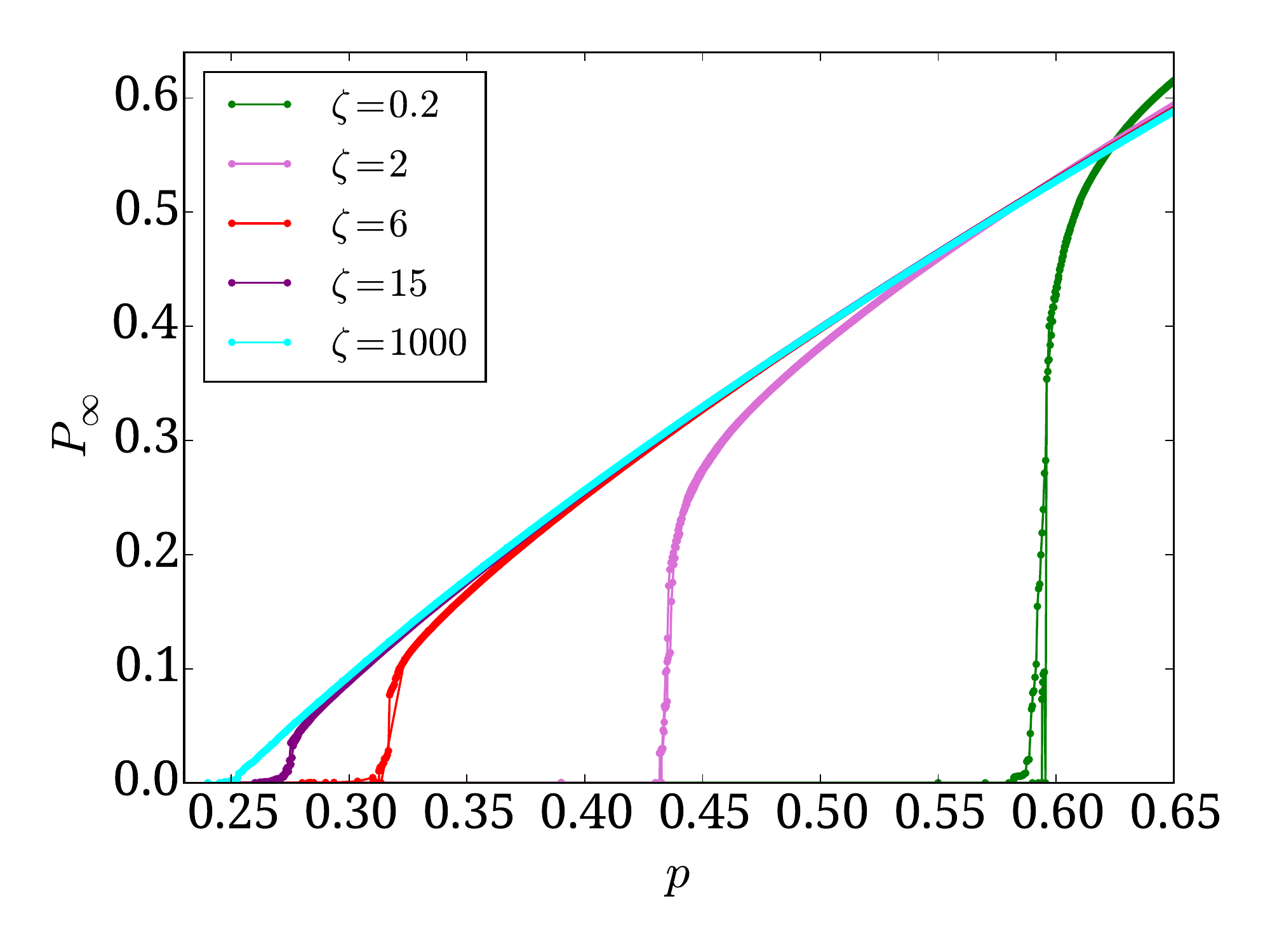}\label{subfig:single_perc}}
\subfloat[multiplex]{\includegraphics[width=0.4\textwidth]{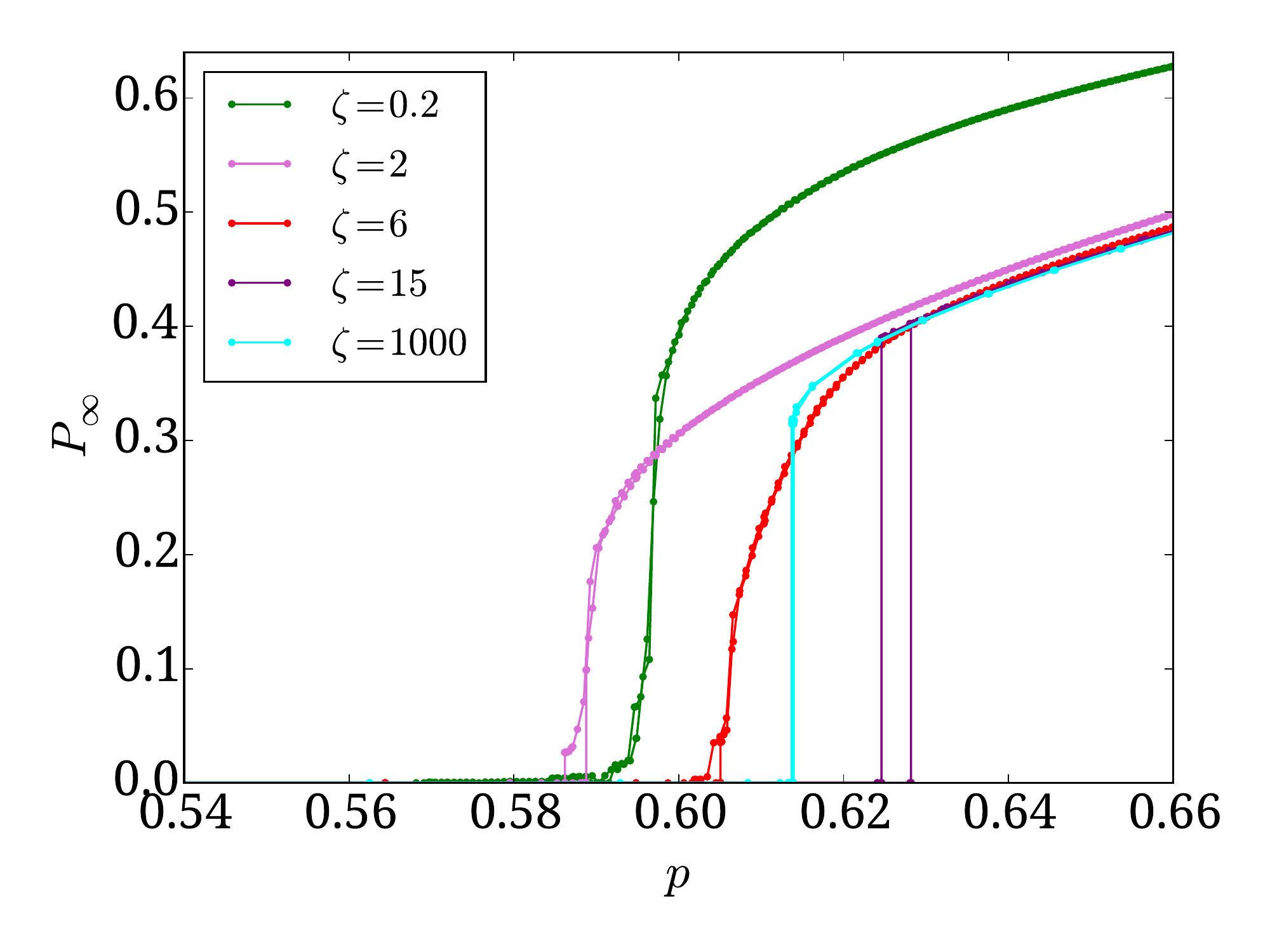}\label{subfig:multi_perc}}
\caption{
\textbf{Percolation of \textbf{(a)} single-layer  and \textbf{(b)} multiplex networks with links of characteristic geographic length.}
The fraction of the network in the largest connected component ($P_\infty$) as a function of the fraction of nodes remaining after random removal of a fraction $1-p$ nodes.  
 \textbf{(a)} In single networks, the transition is always second-order and the critical behavior above $p_c$ is the same for all finite values of $\zeta$ ($\zeta\ll L$).
 The value of $p_c$ decreases quickly and monotonically with $\zeta$ ($\zeta>1$). 
 \textbf{(b)} In multiplex networks, the transition is comparable to single networks for $\zeta = 0.2$ ($p_c \approx 0.5927$) but $p_c$ \textit{increases} as $\zeta$ increases.
 This continues until the maximal value is reached (at $\zeta_c$) and the transition becomes first-order.
 The case of $\zeta = 1000$ is very close to purely random and has $p_c \approx 2.4554 / \kk = 0.61385$.
 Each line represents an individual realization for $\kk = 4$ and $L = 4000$.}
\label{fig:percolation}
\end{figure*}
\section{Model}

To model connectivity links of characteristic length in each layer, we construct the network as follows.
We begin by assigning each node an $(x,y)$  coordinate with integers $x,y\in [0,1,\ldots,L)$.
To construct the links in each layer, we select a source node at random with coordinates $(x_0,y_0)$ and draw a length $l$ according to $P(l) \sim e^{-l/\zeta}$.
We choose the permitted link length $(dx,dy)$ which is closest to fulfilling $l = \sqrt{dx^2 + dy^2}$,  
select one of the eight length-preserving permutations $(dx\leftrightarrow-dx, dy\leftrightarrow-dy, dx\leftrightarrow dy)$ uniformly at random 
and make a link to node $(x_1,y_1)$ with $x_1 = x_0 + dx$ , $y_1 = y_0 + dy$.
This process is executed independently in each layer and is continued until the desired number of links ($N\kk/2$) is obtained.
For simplicity, we use the same characteristic length $\zeta$ and average degree $\kk$ in each layer.
However, because they are constructed independently, the links in each layer are different (as demonstrated in Fig. \ref{fig:layers} and Fig. \ref{fig:overlap}) 
and this disorder enables the critical behavior which we describe below.

We then perform site percolation by removing a fraction $1-p$ of the nodes from the system and finding the mutual giant component \cite{buldyrev-nature2010}.
When a node is removed, it causes damage to nodes in both layers due to the connectivity links in each layer, which are also removed.
However, since the connectivity links are not the same in both layers, there will be nodes that are connected to the giant component in one layer but are disconnected in the other layer.
Since the node functionality requires connectivity in \textit{both} layers, such nodes will fail, causing further damage in the system.
This leads to the cascading failures as demonstrated in Figs. \ref{fig:cascade-diagram} and Fig. \ref{fig:hole}, which are similar to those described in \cite{buldyrev-nature2010,parshani-epl2010,wei-prl2012,hu-pre2013,danziger-jcomnets2014,dong-pre2014,berezin-scireports2015}.

\section{Results}
\begin{figure*}
\centering
\hspace*{-5mm}
\subfloat[$\kk = 3$]{\includegraphics[trim = 20 30 24 28, clip=true, width=0.26\textwidth]{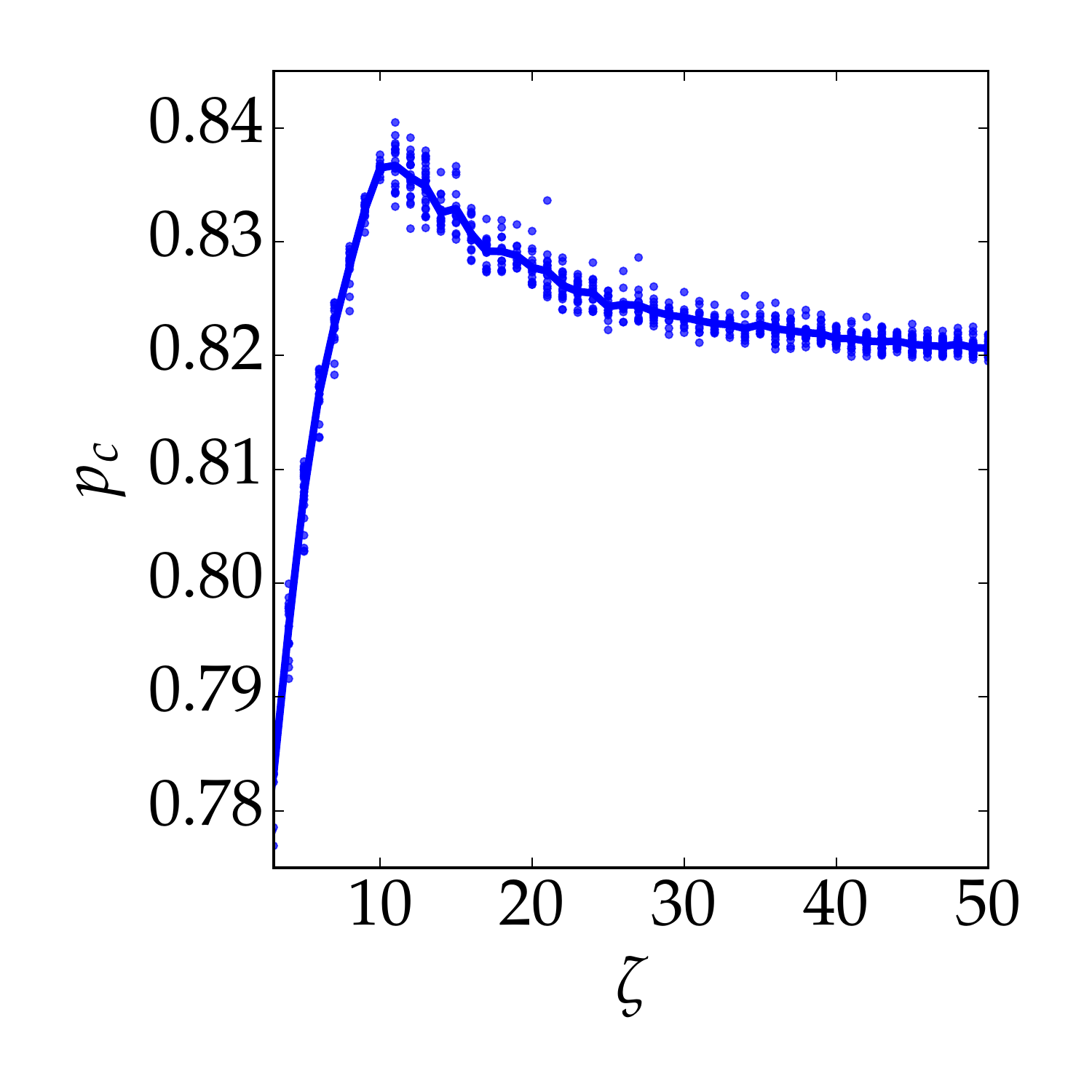}\label{subfig:pck3}}
\subfloat[$\kk = 4$]{\includegraphics[trim = 20 30 24 28, clip=true, width=0.26\textwidth]{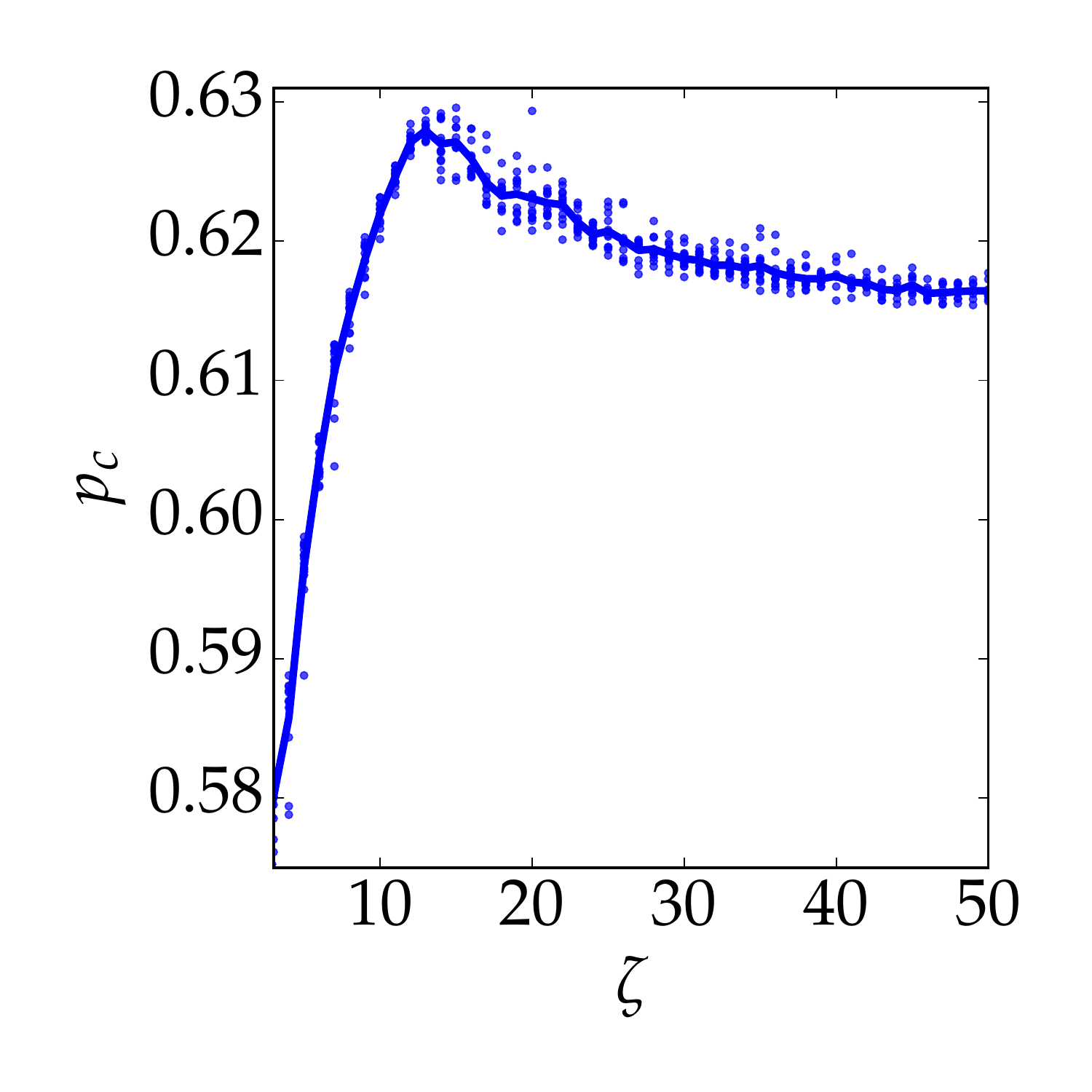}\label{subfig:pck4}}
\subfloat[$\kk = 3$]{\includegraphics[trim = 20 30 24 28, clip=true, width=0.26\textwidth]{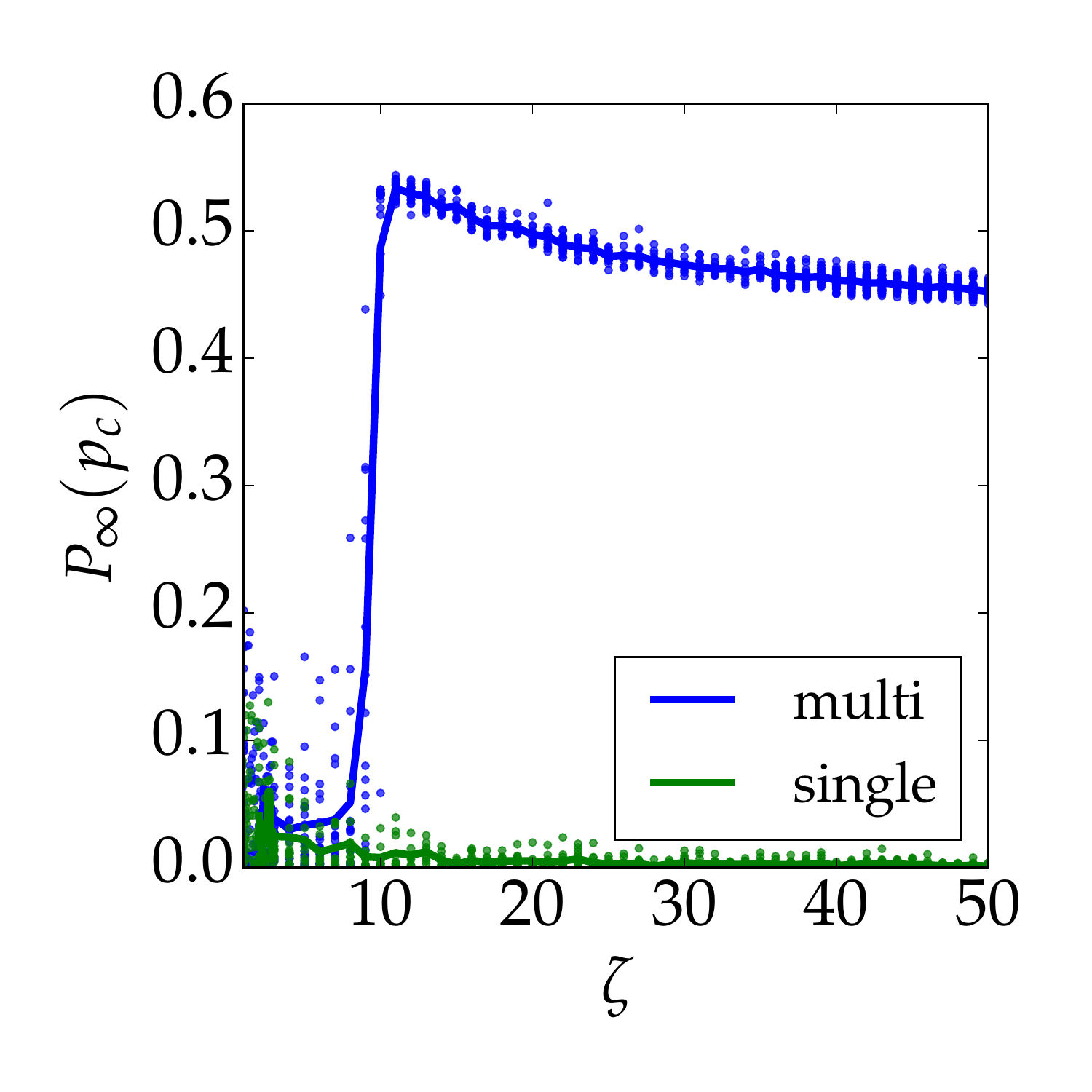}\label{subfig:njumpk3}}
\subfloat[$\kk = 4$]{\includegraphics[trim = 20 30 24 28, clip=true, width=0.26\textwidth]{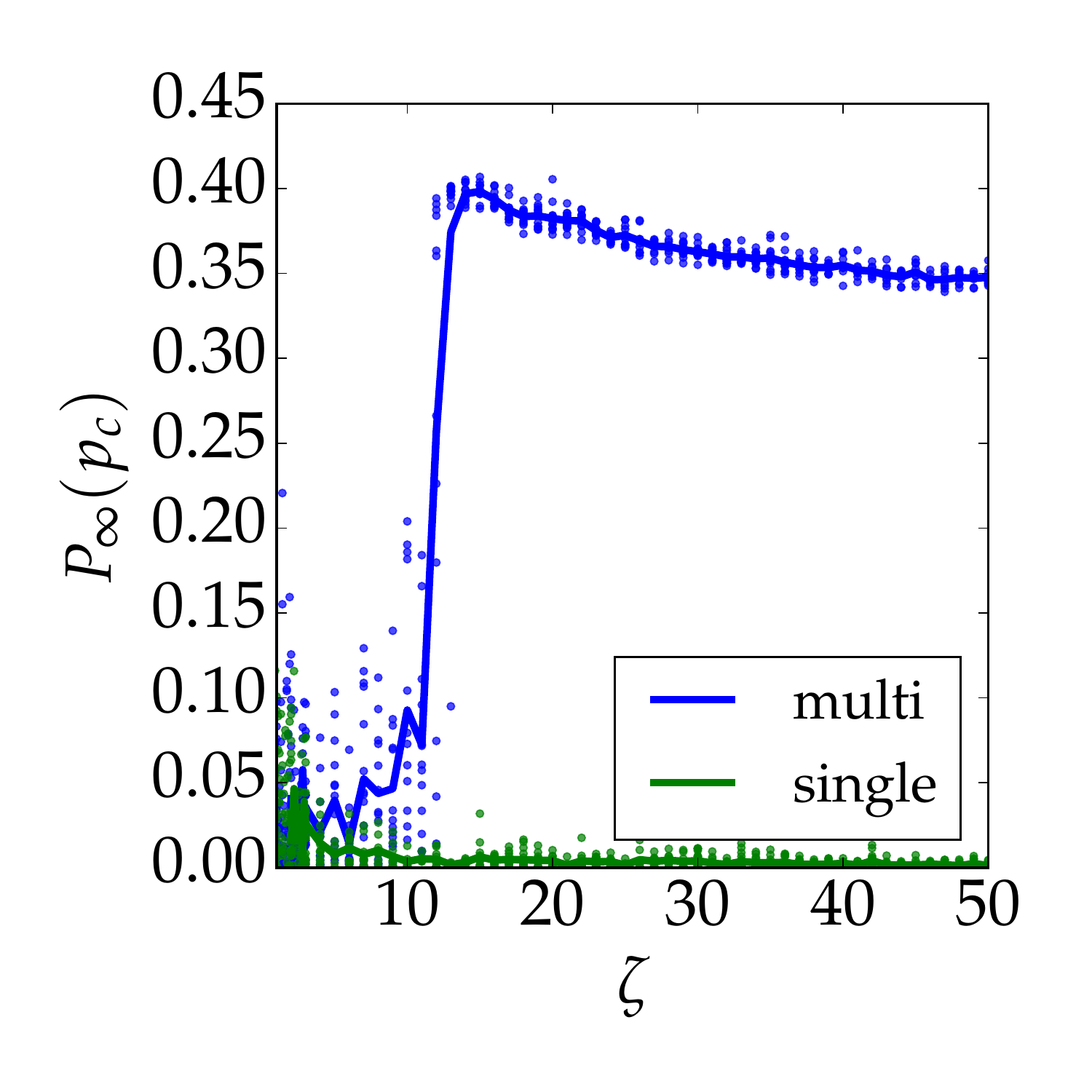}\label{subfig:njumpk4}}
\caption{
\textbf{The effect of the characteristic link length $\zeta$ on the percolation threshold  and size of the giant component at criticality.}
\textbf{(a-b)} The percolation threshold in multiplex networks increases until it reaches a peak at $\zeta_c$ and then decreases slightly to its asymptotic value of $2.4554/\kk$.
\textbf{(c-d)} The size of the order parameter $P_\infty$ at $p_c$.
Single-layer networks always have an order parameter close to zero at $p_c$, which indicates a second-order transition.
Multiplex networks have a second-order transition for low values of $\zeta$ but at $\zeta_c$ this jumps to a large fraction of the system size, indicating a first-order transition.
All panels have $L=4000$ and are plotted based on at least 10 realizations of each system.
The averages and raw data are plotted for all systems.
\label{fig:pc_Njump}}
\end{figure*}

Since we are not aware of a discussion of the percolation properties of this topology for single networks, we briefly describe those properties here and in the Appendix.
In the limit of $\zeta \rightarrow 0$, the only permitted links will be to nearest neighbors (because links of length $<1$ are not accessible) and a square lattice is recovered.
As such, in the case of $\kk = 4$, we recover the standard 2-dimensional percolation behavior with $p_c \approx 0.5927$ \cite{bunde1991fractals,ziff-prl1992} (Fig. \ref{subfig:single_perc}).
When slightly longer (i.e., next nearest neighbor) links are allowed, the system becomes slightly less robust as discussed in the Appendix.
As $\zeta$ increases further, the robustness increases and in the limit $\zeta \rightarrow \infty$, all lengths are equally likely to be drawn and the topology reverts to purely random (\er~topology) with $p_c = 1 / \kk = 0.25$ (see Fig. \ref{subfig:single_perc}.)
Thus, similar to rewiring probability in the original small world model \cite{watts-nature1998}, we have a single parameter $\zeta$ which allows us to smoothly transition from lattice to random topology.
For all values of $\zeta$, a single network undergoes a second-order transition (Fig. \ref{subfig:njumpk3} and \ref{subfig:njumpk4}).


\begin{figure}
\subfloat[$t=30$]{\includegraphics[width=0.47\linewidth]{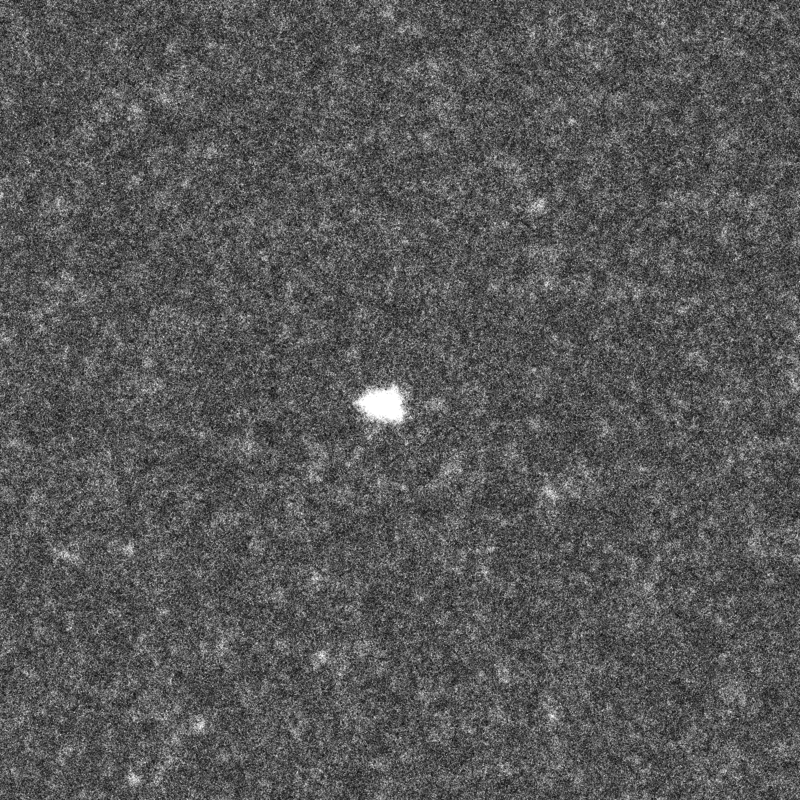}}\hfill
\subfloat[$t=100$]{\includegraphics[width=0.47\linewidth]{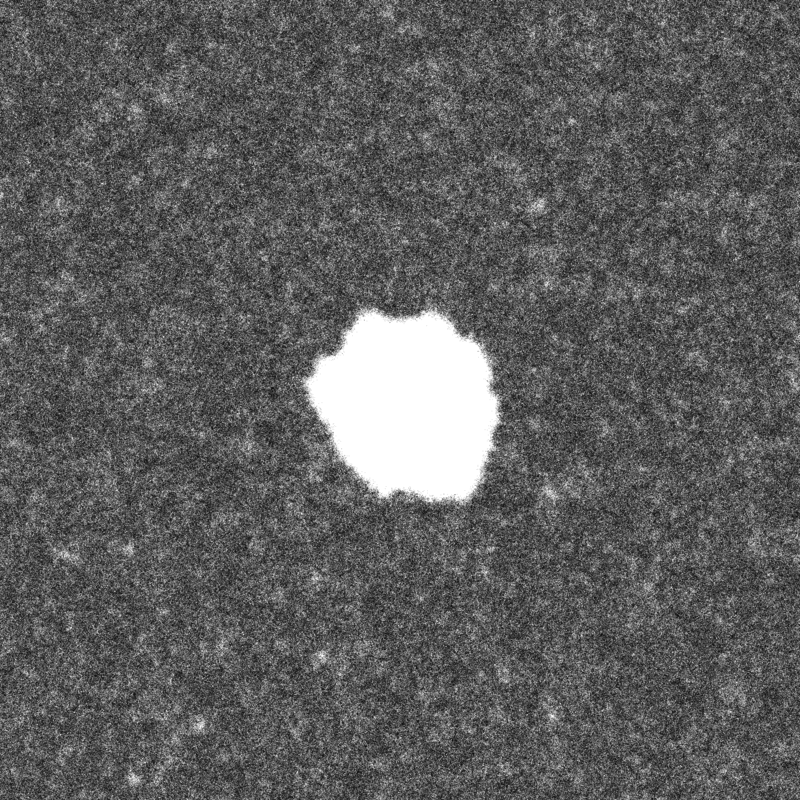}}\\
\subfloat[$t=200$]{\includegraphics[width=0.47\linewidth]{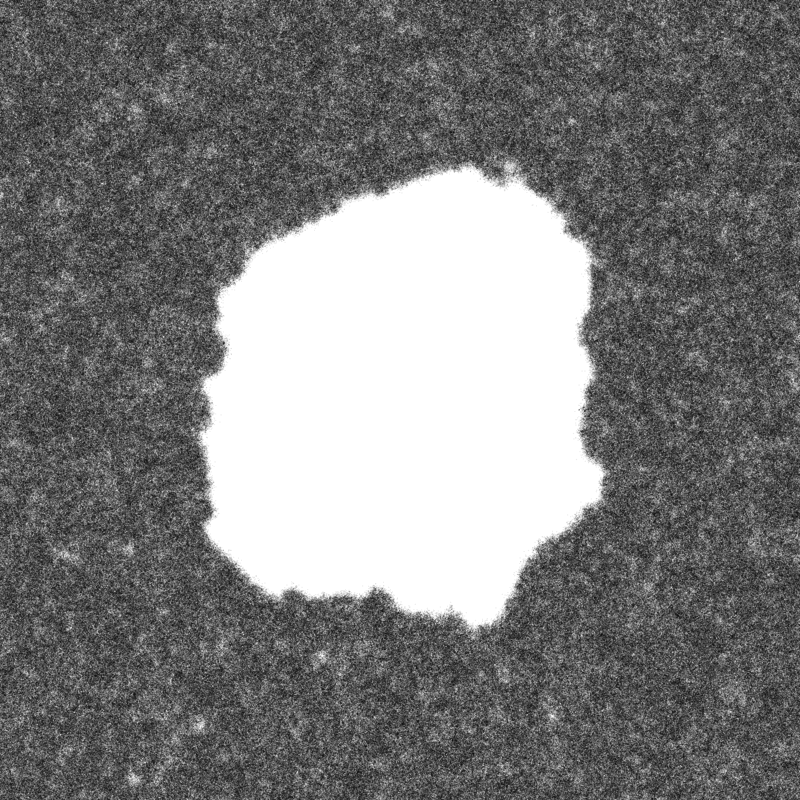}}\hfill
\subfloat[$t=300$]{\includegraphics[width=0.47\linewidth]{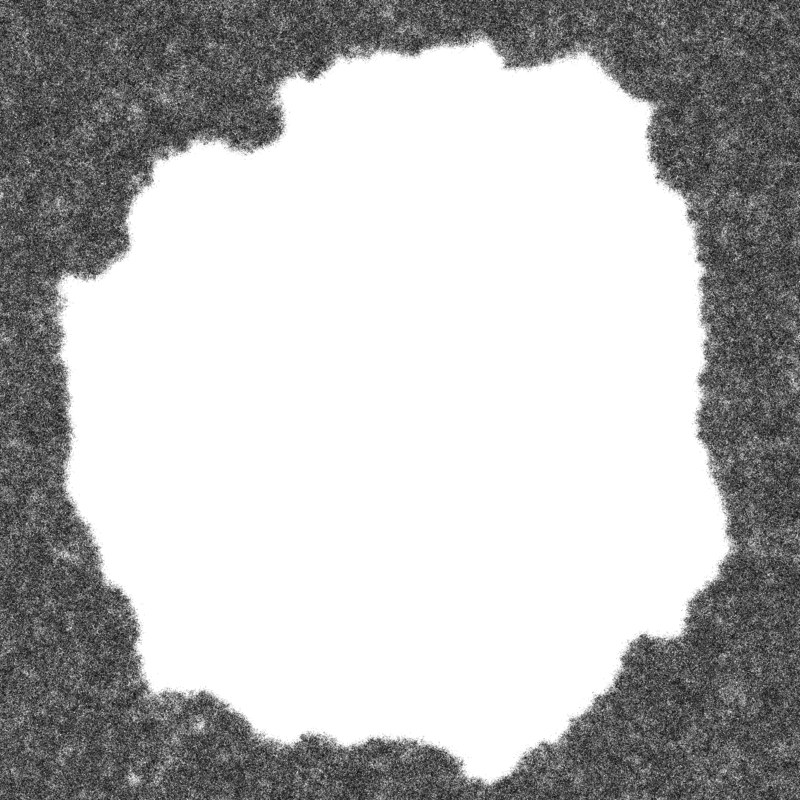}}
\caption{\textbf{Dynamic evolution of cascading failures.}
Here we show the nodes of the multiplex network, colored black if functional and white if not.
After the emergence of a large enough hole (due to random fluctuations), the damage is relayed predominantly to the area around the edge of the hole (due to the finite link lengths) which leads to the nodes around the interface becoming disconnected and propagation of the damage until the system disintegrates.
Similar dynamics have been observed in interdependent lattices with dependency lengths of finite length \cite{wei-prl2012,danziger-jcomnets2014}.
In this figure $L=4000$, $\zeta=15$, $\kk=4$.}\label{fig:hole}
\end{figure}

In multiplex networks, where connectivity to the giant component in both layers is required, cascading failures emerge \cite{buldyrev-nature2010,baxter-prl2012,peixoto-prl2012,danziger-ndes2014}.
For $\zeta\approx 0$, large cascading failures do not emerge.
This is because the multi-layer structure is mostly redundant and the difference between connectivity in one layer or both layers is negligible.
However, once $\zeta$ becomes long enough (above $\zeta_c$), intensive cascading failures emerge and the system undergoes an abrupt, first-order transition, similar to the transition in interdependent lattices \cite{wei-prl2012}, as shown in Fig. \ref{fig:pc_Njump} and \ref{fig:hole}.
As $\zeta$ becomes even longer, $p_c$ decreases and slowly approaches its asymptotic value of $2.4554/\kk$ as known from interdependent \er~ networks \cite{buldyrev-nature2010}, (Fig. \ref{subfig:pck3} and \ref{subfig:pck4}).

In interdependent lattices the mechanism of the first-order transition for dependency links of large finite length is a propagating spinodal interface \cite{wei-prl2012,danziger-sitis2013,danziger-jcomnets2014,berezin-scireports2015}.
On a microscopic level, the first-order transition takes place by the emergence of a hole due to random fluctuations of characteristic length $\xi$ (the percolation correlation length) which then propagates through the system.
This propagation is enabled by the cascade dynamics (cf. Fig. \ref{fig:cascade-diagram}) and dependency links which relay the damage caused by the hole into a concentrated area around the hole's edge.
It would seem that in order for this phenomenon to take place, long dependency links are needed and longer connectivity link lengths would be insufficient to sustain damage propagation.
The reason that the dependency links have a stronger influence on robustness is that in order for a node to fail due to connectivity, all of its links to the giant component need to fail whereas with dependency, an otherwise well-connected node will fail if the single node that it depends on fails.

We find that, in fact, this is not the case.
Connectivity link lengths which are above a critical length, $\zeta_c$, but much smaller than the total system size are sufficient to cause the cascading failures observed in lattice models, even when the dependency links have zero length.
Indeed, the first order transition lacks scaling behavior just above $p_c$ (Fig. \ref{subfig:multi_perc}) and is characterized by a slow spreading process (Fig. \ref{fig:hole}), just like interdependent lattices described in Refs. \cite{wei-prl2012,danziger-sitis2013,danziger-jcomnets2014}.
However, the maximum $p_c$ in this system is lower than the comparable lattice system: $\approx 0.64$ ($\kk = 4$ here) vs. $\approx 0.74$ (interdependent lattices).
This is because, in order for the connectivity links to effectively relay the damage from the hole, the system must be closer to criticality.
Indeed, only when the average degree of each node is 1 do the connectivity and dependency links have the same effect upon it.

Surprisingly, the highly localized topology which emerges from spatial embedding makes the system more susceptible to cascading failures.
This is due to the fact that, because the damage from an emergent (or induced \cite{berezin-scireports2015}) hole is relayed by the cascade dynamics to the neighborhood of its interface, 
the nodes near the edge of the hole are far more likely to become disconnected.
This is related to work on information diffusion in social networks, where it was shown that high modularity makes viral cascades more likely to occur due to the increased likelihood of multiple exposure to the information \cite{centola-science2010,weng-scireports2013,nematzadeh-prl2014}.

In single layer networks, $p_c$ decreases monotonically as $\zeta$ increases from $\zeta \approx 0.5$ to the limit of $\zeta = \infty$.
In contrast, in multiplex networks, $p_c$ increases until $\zeta_c$ and then decreases monotonically thereafter.
The peak in $p_c(\zeta)$ is due to the fact that the size of the critical hole that is needed to trigger the transition scales linearly with $\zeta$ \cite{berezin-scireports2015} and the size of emergent holes above the percolation threshold, $\xi(p)$, decreases with $p$ \cite{bunde1991fractals}.  
This would indicate that the smaller $\zeta$ is, the smaller the critical hole needs to be and that $p_c$ would increase monotonically as $\zeta$ decreases.
However, when $\zeta < \zeta_c$ there is not enough space between the emergent hole and the extent of the damage propagation ($\zeta$) for the network to disintegrate and the small emergent holes remain in place \cite{wei-prl2012,danziger-sitis2013,danziger-jcomnets2014,berezin-scireports2015}.

For interdependent lattices with dependency links of finite length, the single network case is recovered when the dependency links have length zero ($r=0$) \cite{wei-prl2012}.  
In spatially embedded multiplex networks the same limit is recovered as $\zeta\rightarrow 0$ due to overlapping links.
The fraction of common connectivity links between two interdependent networks or between two layers in a multiplex network is called intersimilarity \cite{parshani-epl2010,hu-pre2013} or overlap \cite{cellai-pre2013,li-newjphysics2013}.
The cascading failures and abrupt transitions which characterize interdependent networks decrease as overlap increases.
In the limit of total overlap, they disappear altogether because the system is composed of two exact copies of the same network and the cascade shown in Fig. \ref{fig:cascade-diagram} does not take place at all.

\begin{figure}
\centering
\includegraphics[width=0.9\linewidth]{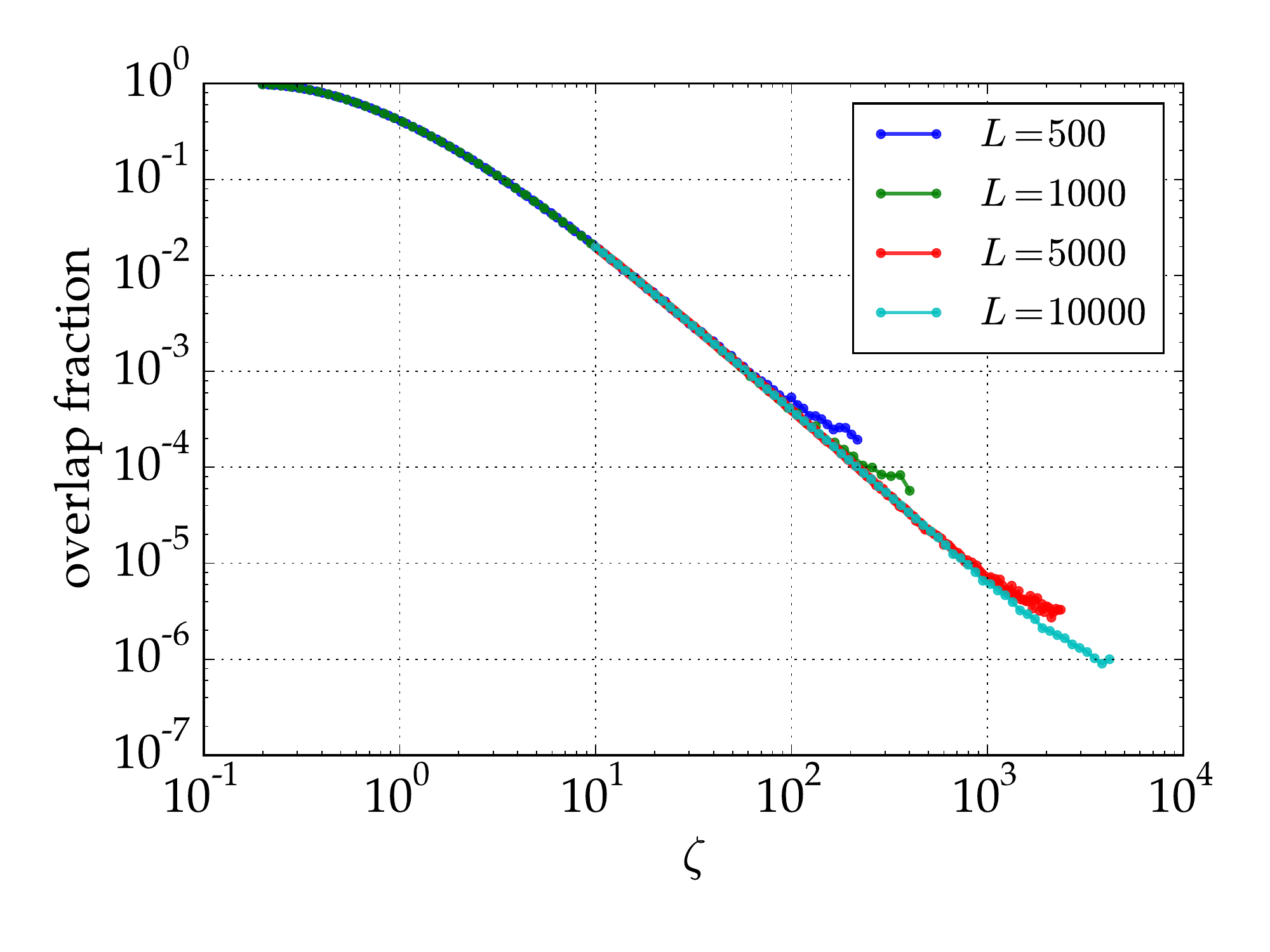}
\caption{\textbf{The fraction of overlapping nodes.}  
The fraction of overlapping nodes is determined as the number of common links across both layers divided by the total number of links in each layer.
When  $\zeta \approx 0$, the overlap is maximal and the networks are identical (for $\kk = 4$, as in this figure).
As $\zeta$ increases, the fraction decreases with an exponent of $\approx -1.8$, see text for discussion.
}
\label{fig:overlap}
\end{figure}

In multiplex networks with links of characteristic length, the extent of the overlap can be estimating by considering the probability that, given the same source node, two links lead to the same target node.  
In the continuum limit this is proportionate to the probability that the links have the same length and the same direction.  
The system is isotropic by construction so the directional condition is simply $1/2\pi r$, the size of a ring of radius $r$.
We obtain
\begin{equation}
P(overlap) \sim \int_1^\infty \frac{P^2(r)}{2\pi r} dr \sim \frac{1}{\zeta^2}  \int_1^\infty \frac{e^{-2r/\zeta}}{2\pi r} dr \sim \frac{1}{\zeta^2}
\end{equation}
We find that the scaling in our system is scale-free, with an exponent of $\approx -1.8$ (Fig. \ref{fig:overlap}).
We hypothesize that the deviation from the continuum calculation is due to the fact that with the discretization of space that we introduce, links that would otherwise have been distinct are unified and the critical exponent is reduced from $-2$ to $\approx -1.8$.

Unlike studies of random multilayer networks with overlap \cite{parshani-epl2010,hu-pre2013,cellai-pre2013,li-newjphysics2013}, decreased overlap alone is not enough to enable the critical behavior observed here.
It is only the combination of the disorder (as indicated by decreased overlap) with the spatially embedded links that enables the distinctive first-order transition which we observe here.
\section{Discussion}
Previous models of spatially embedded interdependent networks \cite{wei-prl2012,bashan-naturephysics2013,danziger-sitis2013,danziger-jcomnets2014,shekhtman-pre2014,berezin-scireports2015} have used two-dimensional lattices with dependency links connecting nodes from one network to the other.
The dependency links were also affected by the spatial embedding via the restriction that they have length of up to $r$, a system parameter.
This model led to many important results, but left several important issues unaddressed.
First, the topology of real-world spatially embedded networks is not lattice-like or even strictly planar and it was not clear that results derived on lattices would accurately describe real-world topologies.
Second, the assumption that dependency links are longer than connectivity links does not correspond with what we would expect from critical infrastructure: It is not reasonable to expect a communications station to get power from a distant power plant and not the one nearest to it.
Here we address these problems by modeling spatially embedded interdependent networks as multiplex networks where the dependency relationship is to the nearest node in the other layer and the connectivity links are of finite characteristic length but not uniform or regular.

We find that the most important features of the lattice-based models are reproduced by our new model: second-order and first-order transitions depending on the link length, a first-order transition characterized by the emergence and spreading of a hole and substantially higher vulnerability compared to single-layer networks.
Furthermore, $p_c$ has a maximal value at the $\zeta$ value ($\zeta_c$) that separates the two types of transitions. 
This validates previous work based on lattices and also shows a new way forward for the modelling of critical infrastructure and other spatially embedded multilayer networks.

\begin{acknowledgments}
We acknowledge the  MULTIPLEX (No. 317532) EU project, the Deutsche Forschungsgemeinschaft (DFG), the Israel Science Foundation, ONR and DTRA for financial support.
We also thank Sergey V. Buldyrev for helpful discussions and comments on the model.
\end{acknowledgments}

\appendix*
\section{Percolation threshold in single networks and in multiplex and single networks for $\zeta<3$}
\begin{figure}
\centering
\subfloat[$\kk = 3$]{\includegraphics[width = 0.5\linewidth]{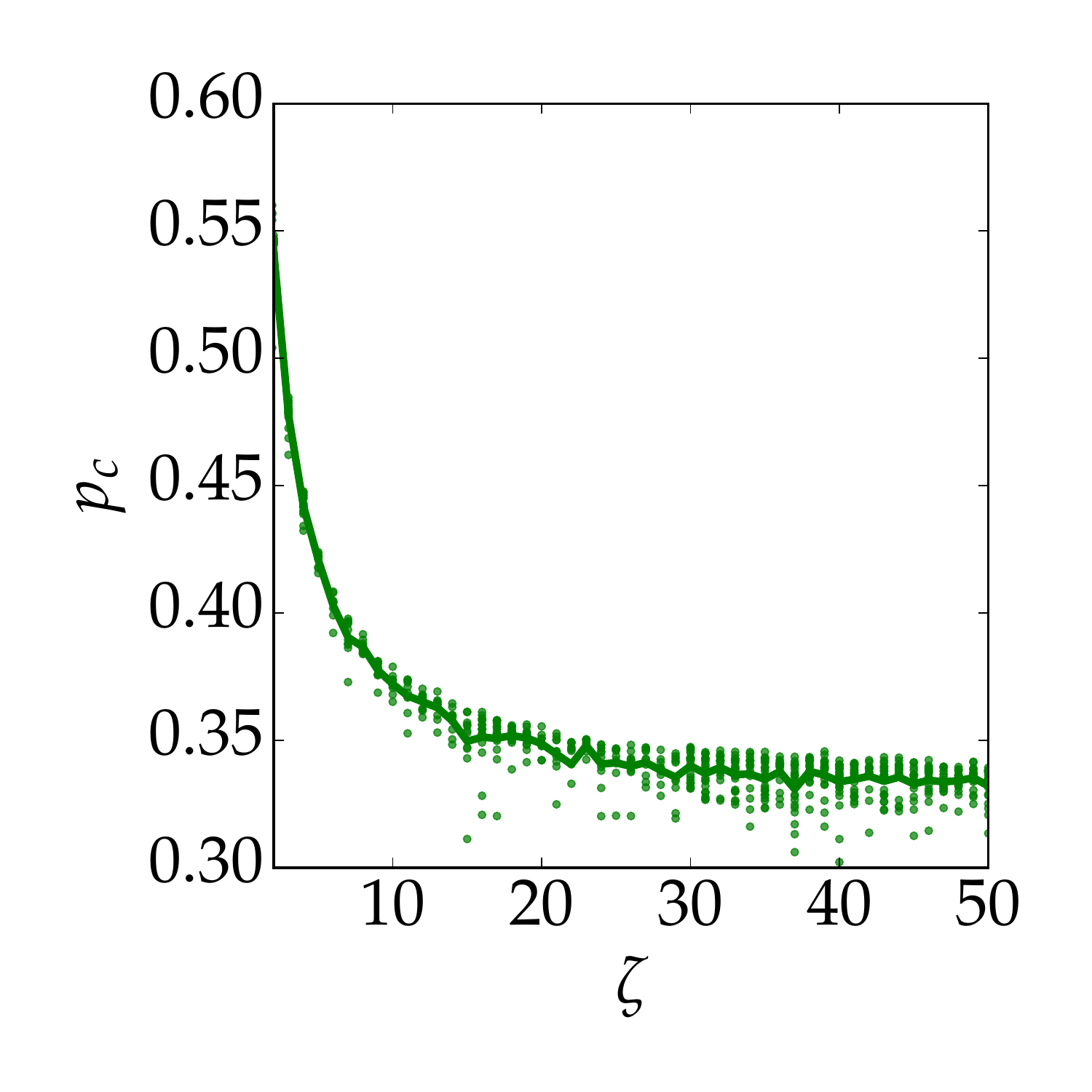}}
\subfloat[$\kk = 4$]{\includegraphics[width = 0.5\linewidth]{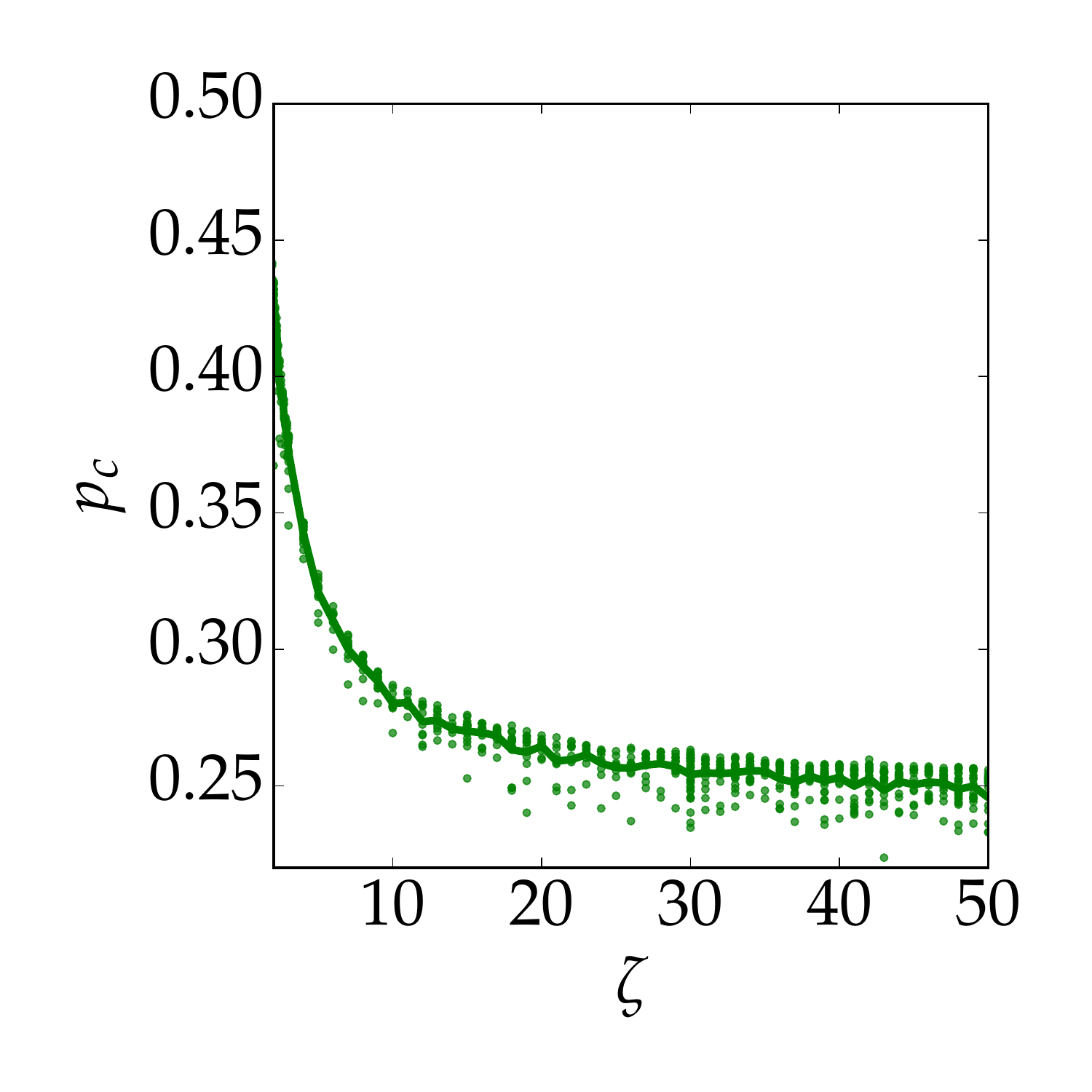}}
\caption{\textbf{Dependence of $p_c$ on $\zeta$ for single networks}
The percolation threshold $p_c$ drops quickly as a function of $\zeta$ and by $\zeta \approx 10$ it is already very close to $p_c = 1/\kk$, the value from \er~networks.
}
\label{fig:pc_zeta_single}
\end{figure}

In single networks, for $\zeta>1$, $p_c$ decreases monotonically with increasing $\zeta$.
This is evident in Fig. \ref{fig:pc_zeta_single}, where $p_c$ is shown to rapidly approach $1/\kk$, the \er~value, as $\zeta$ is increased \cite{newman-book2010, cohen-book2010}.
Thus, we find that the effect of increasing $\zeta$ on single networks is to make them \textit{more} robust, the opposite of its effect in multiplex networks.
\begin{figure}
\centering
\subfloat[$\kk = 3$]{\includegraphics[width=0.5\linewidth]{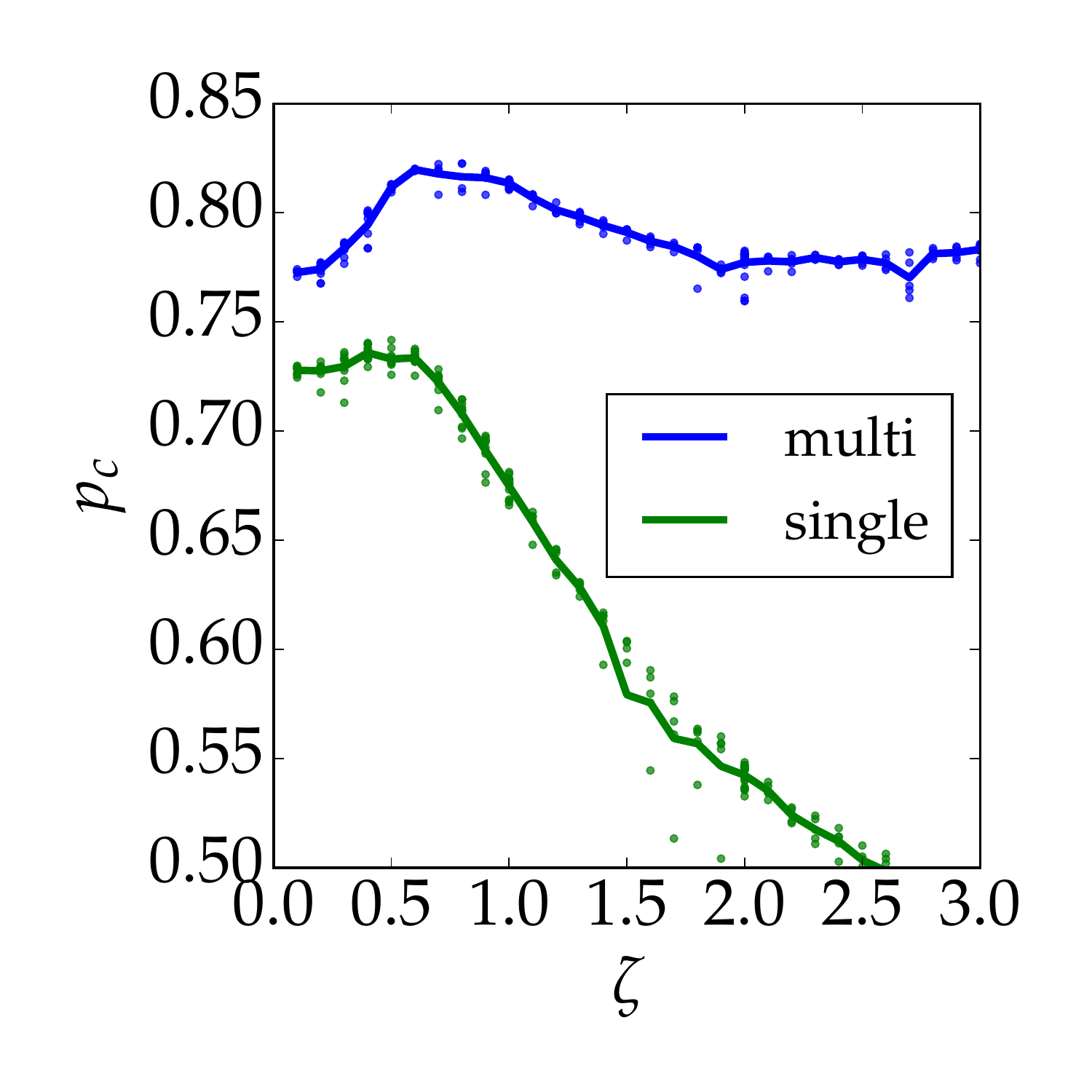}}
\subfloat[$\kk = 4$]{\includegraphics[width=0.5\linewidth]{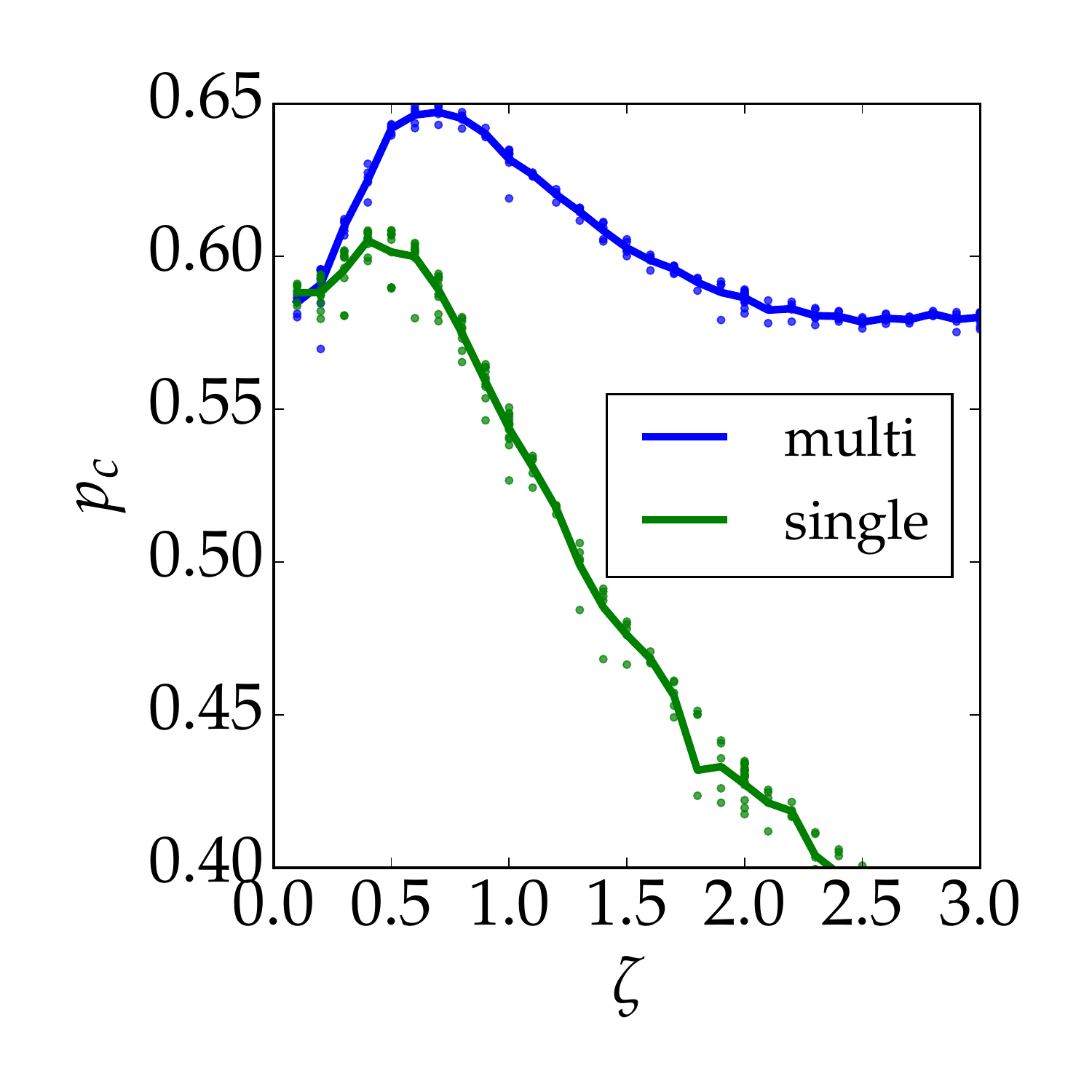}}
\caption{\textbf{Percolation behavior for very short characteristic lengths.}
The percolation threshold increases slightly for all systems with very small $\zeta$.  
For single networks it rapidly falls again but for multiplex networks, $p_c$ remains higher for longer before falling to its $\zeta=0$ level.
This is because in single networks, the only effect is the relative weakness of lattices with complex neighborhoods.
In multiplex networks, this effect is compounded with the weakness engendered by clustering \cite{huang-epl2013,shao-pre2014} and overlap \cite{parshani-epl2010,hu-pre2013,cellai-pre2013,li-newjphysics2013}.
}
\label{fig:shortzeta}
\end{figure}
The behavior of single and multiplex spatially embedded networks is different for $0<\zeta<3$.
Since, this is not the main focus of this study and since it does not affect the critical phenomena, we have avoided discussing it in the main text.
When $\zeta = 0$, only the shortest possible links will be drawn (those with $l = 1$).
Thus, the topology of the network layer will be a pure lattice for $\kk = 4$ and a diluted lattice for $\kk < 4$.
However, for $0<\zeta<3$, some of the nearest neighbor links are exchanged for next nearest neighbor (diagonal) links which are less robust than nearest neighbor only links.
This is a well-known result from bond percolation on lattices with complex neighborhoods, where it was found that $\langle k_c \rangle$ for a next-nearest neighbor lattice ($z=8$) is higher than $\langle k_c \rangle$ for a nearest neighbor lattice ($z=4$) \cite{feng-pre2008}.
For instance, for $\zeta \approx 0.6$ the network layer is a mixture of approximately 85\% nearest neighbor links and 15\% next nearest neighbor links.
As $\zeta$ increases, the links are less redundant and tend to strengthen the network, as seen in Fig. \ref{fig:pc_zeta_single} and Fig. \ref{fig:shortzeta} for single networks.

The robustness is more severely impacted in the multiplex case. This is due to the effects of clustering and disorder.
Combining nearest and next-nearest neighbor links leads to high clustering (number of triangles) which, though having only a minor impact on the robustness of single networks, substantially weakens interdependent networks \cite{huang-epl2013,shao-pre2014}.
Furthermore, the increased disorder decreases the fraction of overlapping links substantially ($\approx 50\%$ as in Fig. \ref{fig:overlap}).
These effects combine in the multiplex case to cause a substantial increase in $p_c$ for $0.3< \zeta <1$.
At $\zeta \approx 1$, the links are no longer redundant, clustering decreases, and the system becomes more stable again. 
However, it is important to note that even though $p_c$ is increased in this interval, the transition is still second-order.
The first-order transition only takes place when the links are substantially longer ($\zeta > \zeta_c$), and the spreading process described above can take place.

\FloatBarrier
\bibliographystyle{apsrev_nourl}
\bibliography{explength}

\begin{thebibliography}{58}
\expandafter\ifx\csname natexlab\endcsname\relax\def\natexlab#1{#1}\fi
\expandafter\ifx\csname bibnamefont\endcsname\relax
  \def\bibnamefont#1{#1}\fi
\expandafter\ifx\csname bibfnamefont\endcsname\relax
  \def\bibfnamefont#1{#1}\fi
\expandafter\ifx\csname citenamefont\endcsname\relax
  \def\citenamefont#1{#1}\fi
\expandafter\ifx\csname url\endcsname\relax
  \def\url#1{\texttt{#1}}\fi
\expandafter\ifx\csname urlprefix\endcsname\relax\def\urlprefix{URL }\fi
\providecommand{\bibinfo}[2]{#2}
\providecommand{\eprint}[2][]{\url{#2}}

\bibitem[{\citenamefont{Doar and Leslie}(1993)}]{doar1993bad}
\bibinfo{author}{\bibfnamefont{M.}~\bibnamefont{Doar}} \bibnamefont{and}
  \bibinfo{author}{\bibfnamefont{I.}~\bibnamefont{Leslie}}, in
  \emph{\bibinfo{booktitle}{INFOCOM'93. Proceedings. Twelfth Annual Joint
  Conference of the IEEE Computer and Communications Societies. Networking:
  Foundation for the Future, IEEE}} (\bibinfo{organization}{IEEE},
  \bibinfo{year}{1993}), pp. \bibinfo{pages}{82--89}.

\bibitem[{\citenamefont{Wei and Estrin}(1993)}]{wei1993comparison}
\bibinfo{author}{\bibfnamefont{L.}~\bibnamefont{Wei}} \bibnamefont{and}
  \bibinfo{author}{\bibfnamefont{D.}~\bibnamefont{Estrin}},
  \bibinfo{journal}{Submitted to INFOCOM} \textbf{\bibinfo{volume}{94}}
  (\bibinfo{year}{1993}).

\bibitem[{\citenamefont{Zegura et~al.}(1997)\citenamefont{Zegura, Calvert, and
  Donahoo}}]{zegura1997quantitative}
\bibinfo{author}{\bibfnamefont{E.~W.} \bibnamefont{Zegura}},
  \bibinfo{author}{\bibfnamefont{K.~L.} \bibnamefont{Calvert}},
  \bibnamefont{and} \bibinfo{author}{\bibfnamefont{M.~J.}
  \bibnamefont{Donahoo}}, \bibinfo{journal}{IEEE/ACM Transactions on Networking
  (TON)} \textbf{\bibinfo{volume}{5}}, \bibinfo{pages}{770}
  (\bibinfo{year}{1997}).

\bibitem[{\citenamefont{Watts and Strogatz}(1998)}]{watts-nature1998}
\bibinfo{author}{\bibfnamefont{D.~J.} \bibnamefont{Watts}} \bibnamefont{and}
  \bibinfo{author}{\bibfnamefont{S.~H.} \bibnamefont{Strogatz}},
  \bibinfo{journal}{Nature} \textbf{\bibinfo{volume}{393}},
  \bibinfo{pages}{440} (\bibinfo{year}{1998}).

\bibitem[{\citenamefont{Penrose}(2003)}]{penrose2003random}
\bibinfo{author}{\bibfnamefont{M.}~\bibnamefont{Penrose}},
  \emph{\bibinfo{title}{Random geometric graphs}}, vol.~\bibinfo{volume}{5}
  (\bibinfo{publisher}{Oxford University Press Oxford}, \bibinfo{year}{2003}).

\bibitem[{\citenamefont{Kleinberg}(2000)}]{kleinberg2000small}
\bibinfo{author}{\bibfnamefont{J.}~\bibnamefont{Kleinberg}}, in
  \emph{\bibinfo{booktitle}{Proceedings of the thirty-second annual ACM
  symposium on Theory of computing}} (\bibinfo{organization}{ACM},
  \bibinfo{year}{2000}), pp. \bibinfo{pages}{163--170}.

\bibitem[{\citenamefont{Kosmidis et~al.}(2008)\citenamefont{Kosmidis, Havlin,
  and Bunde}}]{kosmidis-epl2008}
\bibinfo{author}{\bibfnamefont{K.}~\bibnamefont{Kosmidis}},
  \bibinfo{author}{\bibfnamefont{S.}~\bibnamefont{Havlin}}, \bibnamefont{and}
  \bibinfo{author}{\bibfnamefont{A.}~\bibnamefont{Bunde}},
  \bibinfo{journal}{EPL (Europhysics Letters)} \textbf{\bibinfo{volume}{82}},
  \bibinfo{pages}{48005} (\bibinfo{year}{2008}).

\bibitem[{\citenamefont{Li et~al.}(2011{\natexlab{a}})\citenamefont{Li, Li,
  Kosmidis, Stanley, Bunde, and Havlin}}]{li-epl2011}
\bibinfo{author}{\bibfnamefont{D.}~\bibnamefont{Li}},
  \bibinfo{author}{\bibfnamefont{G.}~\bibnamefont{Li}},
  \bibinfo{author}{\bibfnamefont{K.}~\bibnamefont{Kosmidis}},
  \bibinfo{author}{\bibfnamefont{H.~E.} \bibnamefont{Stanley}},
  \bibinfo{author}{\bibfnamefont{A.}~\bibnamefont{Bunde}}, \bibnamefont{and}
  \bibinfo{author}{\bibfnamefont{S.}~\bibnamefont{Havlin}},
  \bibinfo{journal}{EPL (Europhysics Letters)} \textbf{\bibinfo{volume}{93}},
  \bibinfo{pages}{68004} (\bibinfo{year}{2011}{\natexlab{a}}).

\bibitem[{\citenamefont{Barth{\'e}l\'emy}(2011)}]{barthelemy-physicsreports2011}
\bibinfo{author}{\bibfnamefont{M.}~\bibnamefont{Barth{\'e}l\'emy}},
  \bibinfo{journal}{Physics Reports} \textbf{\bibinfo{volume}{499}},
  \bibinfo{pages}{1 } (\bibinfo{year}{2011}).

\bibitem[{\citenamefont{McAndrew et~al.}(2015)\citenamefont{McAndrew, Danforth,
  and Bagrow}}]{mcandrew2015robustness}
\bibinfo{author}{\bibfnamefont{T.~C.} \bibnamefont{McAndrew}},
  \bibinfo{author}{\bibfnamefont{C.~M.} \bibnamefont{Danforth}},
  \bibnamefont{and} \bibinfo{author}{\bibfnamefont{J.~P.}
  \bibnamefont{Bagrow}}, \bibinfo{journal}{arXiv preprint arXiv:1501.05976}
  (\bibinfo{year}{2015}).

\bibitem[{\citenamefont{Rozenfeld et~al.}(2002)\citenamefont{Rozenfeld, Cohen,
  ben Avraham, and Havlin}}]{rozenfeld-prl2002}
\bibinfo{author}{\bibfnamefont{A.~F.} \bibnamefont{Rozenfeld}},
  \bibinfo{author}{\bibfnamefont{R.}~\bibnamefont{Cohen}},
  \bibinfo{author}{\bibfnamefont{D.}~\bibnamefont{ben Avraham}},
  \bibnamefont{and} \bibinfo{author}{\bibfnamefont{S.}~\bibnamefont{Havlin}},
  \bibinfo{journal}{Phys. Rev. Lett.} \textbf{\bibinfo{volume}{89}},
  \bibinfo{pages}{218701} (\bibinfo{year}{2002}).

\bibitem[{\citenamefont{Bradonji{\'c} et~al.}(2007)\citenamefont{Bradonji{\'c},
  Hagberg, and Percus}}]{bradonjic2007giant}
\bibinfo{author}{\bibfnamefont{M.}~\bibnamefont{Bradonji{\'c}}},
  \bibinfo{author}{\bibfnamefont{A.}~\bibnamefont{Hagberg}}, \bibnamefont{and}
  \bibinfo{author}{\bibfnamefont{A.~G.} \bibnamefont{Percus}}, in
  \emph{\bibinfo{booktitle}{Algorithms and Models for the Web-Graph}}
  (\bibinfo{publisher}{Springer}, \bibinfo{year}{2007}), pp.
  \bibinfo{pages}{209--216}.

\bibitem[{\citenamefont{Hines et~al.}(2010)\citenamefont{Hines, Blumsack,
  Cotilla~Sanchez, and Barrows}}]{hines2010topological}
\bibinfo{author}{\bibfnamefont{P.}~\bibnamefont{Hines}},
  \bibinfo{author}{\bibfnamefont{S.}~\bibnamefont{Blumsack}},
  \bibinfo{author}{\bibfnamefont{E.}~\bibnamefont{Cotilla~Sanchez}},
  \bibnamefont{and} \bibinfo{author}{\bibfnamefont{C.}~\bibnamefont{Barrows}},
  in \emph{\bibinfo{booktitle}{System Sciences (HICSS), 2010 43rd Hawaii
  International Conference on}} (\bibinfo{organization}{IEEE},
  \bibinfo{year}{2010}), pp. \bibinfo{pages}{1--10}.

\bibitem[{\citenamefont{Deka and Vishwanath}(2013)}]{deka-sitis2013}
\bibinfo{author}{\bibfnamefont{D.}~\bibnamefont{Deka}} \bibnamefont{and}
  \bibinfo{author}{\bibfnamefont{S.}~\bibnamefont{Vishwanath}}, in
  \emph{\bibinfo{booktitle}{2013 International Conference on Signal-Image
  Technology \& Internet-Based Systems}} (\bibinfo{year}{2013}), pp.
  \bibinfo{pages}{591--598}, ISBN \bibinfo{isbn}{978-1-4799-3211-5/13}.

\bibitem[{\citenamefont{Manna and Kabakçioglu}(2003)}]{manna-jphysicsa2003}
\bibinfo{author}{\bibfnamefont{S.~S.} \bibnamefont{Manna}} \bibnamefont{and}
  \bibinfo{author}{\bibfnamefont{A.}~\bibnamefont{Kabakçioglu}},
  \bibinfo{journal}{Journal of Physics A: Mathematical and General}
  \textbf{\bibinfo{volume}{36}}, \bibinfo{pages}{L279} (\bibinfo{year}{2003}).

\bibitem[{\citenamefont{Gastner and Newman}(2006)}]{gastner2006spatial}
\bibinfo{author}{\bibfnamefont{M.~T.} \bibnamefont{Gastner}} \bibnamefont{and}
  \bibinfo{author}{\bibfnamefont{M.~E.} \bibnamefont{Newman}},
  \bibinfo{journal}{The European Physical Journal B-Condensed Matter and
  Complex Systems} \textbf{\bibinfo{volume}{49}}, \bibinfo{pages}{247}
  (\bibinfo{year}{2006}).

\bibitem[{\citenamefont{Emmerich et~al.}(2014)\citenamefont{Emmerich, Bunde,
  and Havlin}}]{emmerich2014structural}
\bibinfo{author}{\bibfnamefont{T.}~\bibnamefont{Emmerich}},
  \bibinfo{author}{\bibfnamefont{A.}~\bibnamefont{Bunde}}, \bibnamefont{and}
  \bibinfo{author}{\bibfnamefont{S.}~\bibnamefont{Havlin}},
  \bibinfo{journal}{Physical Review E} \textbf{\bibinfo{volume}{89}},
  \bibinfo{pages}{062806} (\bibinfo{year}{2014}).

\bibitem[{\citenamefont{Ren et~al.}(2014)\citenamefont{Ren, Ercsey-Ravasz,
  Wang, Gonz\'alez, and Toroczkai}}]{ren-naturecomm2014}
\bibinfo{author}{\bibfnamefont{Y.}~\bibnamefont{Ren}},
  \bibinfo{author}{\bibfnamefont{M.}~\bibnamefont{Ercsey-Ravasz}},
  \bibinfo{author}{\bibfnamefont{P.}~\bibnamefont{Wang}},
  \bibinfo{author}{\bibfnamefont{M.~C.} \bibnamefont{Gonz\'alez}},
  \bibnamefont{and}
  \bibinfo{author}{\bibfnamefont{Z.}~\bibnamefont{Toroczkai}},
  \bibinfo{journal}{Nature Communications} \textbf{\bibinfo{volume}{5}},
  \bibinfo{pages}{5347} (\bibinfo{year}{2014}), ISSN \bibinfo{issn}{2041-1723}.

\bibitem[{\citenamefont{Wang et~al.}(2008)\citenamefont{Wang, Thomas, and
  Scaglione}}]{wang-proceedings2008}
\bibinfo{author}{\bibfnamefont{Z.}~\bibnamefont{Wang}},
  \bibinfo{author}{\bibfnamefont{R.~J.} \bibnamefont{Thomas}},
  \bibnamefont{and}
  \bibinfo{author}{\bibfnamefont{A.}~\bibnamefont{Scaglione}},
  \emph{\bibinfo{title}{{Generating Random Topology Power Grids}}}
  (\bibinfo{publisher}{Institute of Electrical and Electronics Engineers},
  \bibinfo{year}{2008}), pp. \bibinfo{pages}{183--183}, ISBN
  \bibinfo{isbn}{0-7695-3075-8}.

\bibitem[{\citenamefont{Grassberger}(2013)}]{grassberger2013sir}
\bibinfo{author}{\bibfnamefont{P.}~\bibnamefont{Grassberger}},
  \bibinfo{journal}{Journal of Statistical Mechanics: Theory and Experiment}
  \textbf{\bibinfo{volume}{2013}}, \bibinfo{pages}{P04004}
  (\bibinfo{year}{2013}).

\bibitem[{\citenamefont{Waxman}(1988)}]{waxman1988routing}
\bibinfo{author}{\bibfnamefont{B.~M.} \bibnamefont{Waxman}},
  \bibinfo{journal}{Selected Areas in Communications, IEEE Journal on}
  \textbf{\bibinfo{volume}{6}}, \bibinfo{pages}{1617} (\bibinfo{year}{1988}).

\bibitem[{\citenamefont{Zhou and Bialek}(2005)}]{zhou-ieee2005}
\bibinfo{author}{\bibfnamefont{Q.}~\bibnamefont{Zhou}} \bibnamefont{and}
  \bibinfo{author}{\bibfnamefont{J.}~\bibnamefont{Bialek}},
  \bibinfo{journal}{Power Systems, IEEE Transactions on}
  \textbf{\bibinfo{volume}{20}}, \bibinfo{pages}{782} (\bibinfo{year}{2005}),
  ISSN \bibinfo{issn}{0885-8950}.

\bibitem[{\citenamefont{{National~Land~Information~Division, National Spatial
  Planning and Regional Policy Bureau, MILT of Japan}}(2012)}]{japanrail}
\bibinfo{author}{\bibnamefont{{National~Land~Information~Division, National
  Spatial Planning and Regional Policy Bureau, MILT of Japan}}},
  \emph{\bibinfo{title}{National railway data}},
  \bibinfo{howpublished}{\url{http://nlftp.mlit.go.jp/ksj/gml/datalist/KsjTmplt-N02.html}}
  (\bibinfo{year}{2012}).

\bibitem[{\citenamefont{Buldyrev et~al.}(2010)\citenamefont{Buldyrev, Parshani,
  Paul, Stanley, and Havlin}}]{buldyrev-nature2010}
\bibinfo{author}{\bibfnamefont{S.~V.} \bibnamefont{Buldyrev}},
  \bibinfo{author}{\bibfnamefont{R.}~\bibnamefont{Parshani}},
  \bibinfo{author}{\bibfnamefont{G.}~\bibnamefont{Paul}},
  \bibinfo{author}{\bibfnamefont{H.~E.} \bibnamefont{Stanley}},
  \bibnamefont{and} \bibinfo{author}{\bibfnamefont{S.}~\bibnamefont{Havlin}},
  \bibinfo{journal}{Nature} \textbf{\bibinfo{volume}{464}},
  \bibinfo{pages}{1025} (\bibinfo{year}{2010}).

\bibitem[{\citenamefont{Gao et~al.}(2012)\citenamefont{Gao, Buldyrev, Stanley,
  and Havlin}}]{gao-naturephysics2012}
\bibinfo{author}{\bibfnamefont{J.}~\bibnamefont{Gao}},
  \bibinfo{author}{\bibfnamefont{S.~V.} \bibnamefont{Buldyrev}},
  \bibinfo{author}{\bibfnamefont{H.~E.} \bibnamefont{Stanley}},
  \bibnamefont{and} \bibinfo{author}{\bibfnamefont{S.}~\bibnamefont{Havlin}},
  \bibinfo{journal}{Nature Physics} \textbf{\bibinfo{volume}{8}},
  \bibinfo{pages}{40} (\bibinfo{year}{2012}).

\bibitem[{\citenamefont{Peixoto and Bornholdt}(2012)}]{peixoto-prl2012}
\bibinfo{author}{\bibfnamefont{T.~P.} \bibnamefont{Peixoto}} \bibnamefont{and}
  \bibinfo{author}{\bibfnamefont{S.}~\bibnamefont{Bornholdt}},
  \bibinfo{journal}{Phys. Rev. Lett.} \textbf{\bibinfo{volume}{109}},
  \bibinfo{pages}{118703} (\bibinfo{year}{2012}).

\bibitem[{\citenamefont{Radicchi and
  Arenas}(2013)}]{radicchi-naturephysics2013}
\bibinfo{author}{\bibfnamefont{F.}~\bibnamefont{Radicchi}} \bibnamefont{and}
  \bibinfo{author}{\bibfnamefont{A.}~\bibnamefont{Arenas}},
  \bibinfo{journal}{Nature Physics} \textbf{\bibinfo{volume}{9}},
  \bibinfo{pages}{717} (\bibinfo{year}{2013}).

\bibitem[{\citenamefont{Kivel{\"a} et~al.}(2014)\citenamefont{Kivel{\"a},
  Arenas, Barth\'el\'emy, Gleeson, Moreno, and Porter}}]{kivela-jcomnets2014}
\bibinfo{author}{\bibfnamefont{M.}~\bibnamefont{Kivel{\"a}}},
  \bibinfo{author}{\bibfnamefont{A.}~\bibnamefont{Arenas}},
  \bibinfo{author}{\bibfnamefont{M.}~\bibnamefont{Barth\'el\'emy}},
  \bibinfo{author}{\bibfnamefont{J.~P.} \bibnamefont{Gleeson}},
  \bibinfo{author}{\bibfnamefont{Y.}~\bibnamefont{Moreno}}, \bibnamefont{and}
  \bibinfo{author}{\bibfnamefont{M.~A.} \bibnamefont{Porter}},
  \bibinfo{journal}{Journal of Complex Networks} \textbf{\bibinfo{volume}{2}},
  \bibinfo{pages}{203} (\bibinfo{year}{2014}).

\bibitem[{\citenamefont{Boccaletti et~al.}(2014)\citenamefont{Boccaletti,
  Bianconi, Criado, Del~Genio, G{\'o}mez-Garde{\~n}es, Romance, Sendina-Nadal,
  Wang, and Zanin}}]{boccaletti-physicsreports2014}
\bibinfo{author}{\bibfnamefont{S.}~\bibnamefont{Boccaletti}},
  \bibinfo{author}{\bibfnamefont{G.}~\bibnamefont{Bianconi}},
  \bibinfo{author}{\bibfnamefont{R.}~\bibnamefont{Criado}},
  \bibinfo{author}{\bibfnamefont{C.}~\bibnamefont{Del~Genio}},
  \bibinfo{author}{\bibfnamefont{J.}~\bibnamefont{G{\'o}mez-Garde{\~n}es}},
  \bibinfo{author}{\bibfnamefont{M.}~\bibnamefont{Romance}},
  \bibinfo{author}{\bibfnamefont{I.}~\bibnamefont{Sendina-Nadal}},
  \bibinfo{author}{\bibfnamefont{Z.}~\bibnamefont{Wang}}, \bibnamefont{and}
  \bibinfo{author}{\bibfnamefont{M.}~\bibnamefont{Zanin}},
  \bibinfo{journal}{Physics Reports}  (\bibinfo{year}{2014}), ISSN
  \bibinfo{issn}{0370-1573}.

\bibitem[{\citenamefont{Li et~al.}(2011{\natexlab{b}})\citenamefont{Li,
  Kosmidis, Bunde, and Havlin}}]{li-naturephysics2011}
\bibinfo{author}{\bibfnamefont{D.}~\bibnamefont{Li}},
  \bibinfo{author}{\bibfnamefont{K.}~\bibnamefont{Kosmidis}},
  \bibinfo{author}{\bibfnamefont{A.}~\bibnamefont{Bunde}}, \bibnamefont{and}
  \bibinfo{author}{\bibfnamefont{S.}~\bibnamefont{Havlin}},
  \bibinfo{journal}{Nature Physics} \textbf{\bibinfo{volume}{7}},
  \bibinfo{pages}{481} (\bibinfo{year}{2011}{\natexlab{b}}).

\bibitem[{\citenamefont{Rinaldi et~al.}(2001)\citenamefont{Rinaldi, Peerenboom,
  and Kelly}}]{rinaldi-ieee2001}
\bibinfo{author}{\bibfnamefont{S.}~\bibnamefont{Rinaldi}},
  \bibinfo{author}{\bibfnamefont{J.}~\bibnamefont{Peerenboom}},
  \bibnamefont{and} \bibinfo{author}{\bibfnamefont{T.}~\bibnamefont{Kelly}},
  \bibinfo{journal}{Control Systems, IEEE} \textbf{\bibinfo{volume}{21}},
  \bibinfo{pages}{11} (\bibinfo{year}{2001}), ISSN \bibinfo{issn}{1066-033X}.

\bibitem[{\citenamefont{Hokstad et~al.}(2012)\citenamefont{Hokstad, Utne, and
  Vatn}}]{hokstad-book2012}
\bibinfo{author}{\bibfnamefont{P.}~\bibnamefont{Hokstad}},
  \bibinfo{author}{\bibfnamefont{I.}~\bibnamefont{Utne}}, \bibnamefont{and}
  \bibinfo{author}{\bibfnamefont{J.}~\bibnamefont{Vatn}},
  \emph{\bibinfo{title}{Risk and Interdependencies in Critical Infrastructures:
  A Guideline for Analysis}}, Springer Series in Reliability Engineering
  (\bibinfo{publisher}{Springer}, \bibinfo{year}{2012}), ISBN
  \bibinfo{isbn}{9781447146612}.

\bibitem[{\citenamefont{Helbing}(2013)}]{helbing-nature2013}
\bibinfo{author}{\bibfnamefont{D.}~\bibnamefont{Helbing}},
  \bibinfo{journal}{Nature} \textbf{\bibinfo{volume}{497}},
  \bibinfo{pages}{51–59} (\bibinfo{year}{2013}), ISSN
  \bibinfo{issn}{1476-4687}.

\bibitem[{\citenamefont{Li et~al.}(2012)\citenamefont{Li, Bashan, Buldyrev,
  Stanley, and Havlin}}]{wei-prl2012}
\bibinfo{author}{\bibfnamefont{W.}~\bibnamefont{Li}},
  \bibinfo{author}{\bibfnamefont{A.}~\bibnamefont{Bashan}},
  \bibinfo{author}{\bibfnamefont{S.~V.} \bibnamefont{Buldyrev}},
  \bibinfo{author}{\bibfnamefont{H.~E.} \bibnamefont{Stanley}},
  \bibnamefont{and} \bibinfo{author}{\bibfnamefont{S.}~\bibnamefont{Havlin}},
  \bibinfo{journal}{Phys. Rev. Lett.} \textbf{\bibinfo{volume}{108}},
  \bibinfo{pages}{228702} (\bibinfo{year}{2012}).

\bibitem[{\citenamefont{Danziger et~al.}(2015)\citenamefont{Danziger, Bashan,
  and Havlin}}]{danziger-newjphysics2015}
\bibinfo{author}{\bibfnamefont{M.~M.} \bibnamefont{Danziger}},
  \bibinfo{author}{\bibfnamefont{A.}~\bibnamefont{Bashan}}, \bibnamefont{and}
  \bibinfo{author}{\bibfnamefont{S.}~\bibnamefont{Havlin}},
  \bibinfo{journal}{New Journal of Physics} \textbf{\bibinfo{volume}{17}},
  \bibinfo{pages}{043046} (\bibinfo{year}{2015}).

\bibitem[{\citenamefont{Danziger
  et~al.}(2014{\natexlab{a}})\citenamefont{Danziger, Bashan, Berezin, and
  Havlin}}]{danziger-jcomnets2014}
\bibinfo{author}{\bibfnamefont{M.~M.} \bibnamefont{Danziger}},
  \bibinfo{author}{\bibfnamefont{A.}~\bibnamefont{Bashan}},
  \bibinfo{author}{\bibfnamefont{Y.}~\bibnamefont{Berezin}}, \bibnamefont{and}
  \bibinfo{author}{\bibfnamefont{S.}~\bibnamefont{Havlin}},
  \bibinfo{journal}{Journal of Complex Networks} \textbf{\bibinfo{volume}{2}},
  \bibinfo{pages}{460} (\bibinfo{year}{2014}{\natexlab{a}}).

\bibitem[{\citenamefont{Zhou et~al.}(2014)\citenamefont{Zhou, Bashan, Cohen,
  Berezin, Shnerb, and Havlin}}]{dong-pre2014}
\bibinfo{author}{\bibfnamefont{D.}~\bibnamefont{Zhou}},
  \bibinfo{author}{\bibfnamefont{A.}~\bibnamefont{Bashan}},
  \bibinfo{author}{\bibfnamefont{R.}~\bibnamefont{Cohen}},
  \bibinfo{author}{\bibfnamefont{Y.}~\bibnamefont{Berezin}},
  \bibinfo{author}{\bibfnamefont{N.}~\bibnamefont{Shnerb}}, \bibnamefont{and}
  \bibinfo{author}{\bibfnamefont{S.}~\bibnamefont{Havlin}},
  \bibinfo{journal}{Phys. Rev. E} \textbf{\bibinfo{volume}{90}},
  \bibinfo{pages}{012803} (\bibinfo{year}{2014}).

\bibitem[{\citenamefont{Bashan et~al.}(2013)\citenamefont{Bashan, Berezin,
  Buldyrev, and Havlin}}]{bashan-naturephysics2013}
\bibinfo{author}{\bibfnamefont{A.}~\bibnamefont{Bashan}},
  \bibinfo{author}{\bibfnamefont{Y.}~\bibnamefont{Berezin}},
  \bibinfo{author}{\bibfnamefont{S.~V.} \bibnamefont{Buldyrev}},
  \bibnamefont{and} \bibinfo{author}{\bibfnamefont{S.}~\bibnamefont{Havlin}},
  \bibinfo{journal}{Nature Physics} \textbf{\bibinfo{volume}{9}},
  \bibinfo{pages}{667} (\bibinfo{year}{2013}).

\bibitem[{\citenamefont{Danziger et~al.}(2013)\citenamefont{Danziger, Bashan,
  Berezin, and Havlin}}]{danziger-sitis2013}
\bibinfo{author}{\bibfnamefont{M.~M.} \bibnamefont{Danziger}},
  \bibinfo{author}{\bibfnamefont{A.}~\bibnamefont{Bashan}},
  \bibinfo{author}{\bibfnamefont{Y.}~\bibnamefont{Berezin}}, \bibnamefont{and}
  \bibinfo{author}{\bibfnamefont{S.}~\bibnamefont{Havlin}}, in
  \emph{\bibinfo{booktitle}{Signal-Image Technology Internet-Based Systems
  (SITIS), 2013 International Conference on}} (\bibinfo{year}{2013}), pp.
  \bibinfo{pages}{619--625}.

\bibitem[{\citenamefont{Shekhtman et~al.}(2014)\citenamefont{Shekhtman,
  Berezin, Danziger, and Havlin}}]{shekhtman-pre2014}
\bibinfo{author}{\bibfnamefont{L.~M.} \bibnamefont{Shekhtman}},
  \bibinfo{author}{\bibfnamefont{Y.}~\bibnamefont{Berezin}},
  \bibinfo{author}{\bibfnamefont{M.~M.} \bibnamefont{Danziger}},
  \bibnamefont{and} \bibinfo{author}{\bibfnamefont{S.}~\bibnamefont{Havlin}},
  \bibinfo{journal}{Phys. Rev. E} \textbf{\bibinfo{volume}{90}},
  \bibinfo{pages}{012809} (\bibinfo{year}{2014}).

\bibitem[{\citenamefont{Stippinger and Kert\'esz}(2014)}]{stippinger-physa2014}
\bibinfo{author}{\bibfnamefont{M.}~\bibnamefont{Stippinger}} \bibnamefont{and}
  \bibinfo{author}{\bibfnamefont{J.}~\bibnamefont{Kert\'esz}},
  \bibinfo{journal}{Physica A: Statistical Mechanics and its Applications}
  \textbf{\bibinfo{volume}{416}}, \bibinfo{pages}{481 } (\bibinfo{year}{2014}),
  ISSN \bibinfo{issn}{0378-4371}.

\bibitem[{\citenamefont{Berezin et~al.}(2015)\citenamefont{Berezin, Bashan,
  Danziger, Li, and Havlin}}]{berezin-scireports2015}
\bibinfo{author}{\bibfnamefont{Y.}~\bibnamefont{Berezin}},
  \bibinfo{author}{\bibfnamefont{A.}~\bibnamefont{Bashan}},
  \bibinfo{author}{\bibfnamefont{M.~M.} \bibnamefont{Danziger}},
  \bibinfo{author}{\bibfnamefont{D.}~\bibnamefont{Li}}, \bibnamefont{and}
  \bibinfo{author}{\bibfnamefont{S.}~\bibnamefont{Havlin}},
  \bibinfo{journal}{Scientific Reports} \textbf{\bibinfo{volume}{5}}
  (\bibinfo{year}{2015}).

\bibitem[{\citenamefont{Parshani et~al.}(2010)\citenamefont{Parshani,
  Rozenblat, Ietri, Ducruet, and Havlin}}]{parshani-epl2010}
\bibinfo{author}{\bibfnamefont{R.}~\bibnamefont{Parshani}},
  \bibinfo{author}{\bibfnamefont{C.}~\bibnamefont{Rozenblat}},
  \bibinfo{author}{\bibfnamefont{D.}~\bibnamefont{Ietri}},
  \bibinfo{author}{\bibfnamefont{C.}~\bibnamefont{Ducruet}}, \bibnamefont{and}
  \bibinfo{author}{\bibfnamefont{S.}~\bibnamefont{Havlin}},
  \bibinfo{journal}{EPL (Europhysics Letters)} \textbf{\bibinfo{volume}{92}},
  \bibinfo{pages}{68002} (\bibinfo{year}{2010}).

\bibitem[{\citenamefont{Hu et~al.}(2013)\citenamefont{Hu, Zhou, Zhang, Han,
  Rozenblat, and Havlin}}]{hu-pre2013}
\bibinfo{author}{\bibfnamefont{Y.}~\bibnamefont{Hu}},
  \bibinfo{author}{\bibfnamefont{D.}~\bibnamefont{Zhou}},
  \bibinfo{author}{\bibfnamefont{R.}~\bibnamefont{Zhang}},
  \bibinfo{author}{\bibfnamefont{Z.}~\bibnamefont{Han}},
  \bibinfo{author}{\bibfnamefont{C.}~\bibnamefont{Rozenblat}},
  \bibnamefont{and} \bibinfo{author}{\bibfnamefont{S.}~\bibnamefont{Havlin}},
  \bibinfo{journal}{Phys. Rev. E} \textbf{\bibinfo{volume}{88}},
  \bibinfo{pages}{052805} (\bibinfo{year}{2013}).

\bibitem[{\citenamefont{Bunde and Havlin}(1991)}]{bunde1991fractals}
\bibinfo{author}{\bibfnamefont{A.}~\bibnamefont{Bunde}} \bibnamefont{and}
  \bibinfo{author}{\bibfnamefont{S.}~\bibnamefont{Havlin}},
  \emph{\bibinfo{title}{{Fractals and disordered systems}}}
  (\bibinfo{publisher}{Springer-Verlag New York, Inc.}, \bibinfo{year}{1991}).

\bibitem[{\citenamefont{Ziff}(1992)}]{ziff-prl1992}
\bibinfo{author}{\bibfnamefont{R.}~\bibnamefont{Ziff}}, \bibinfo{journal}{Phys.
  Rev. Lett.} \textbf{\bibinfo{volume}{69}}, \bibinfo{pages}{2670}
  (\bibinfo{year}{1992}).

\bibitem[{\citenamefont{Baxter et~al.}(2012)\citenamefont{Baxter, Dorogovtsev,
  Goltsev, and Mendes}}]{baxter-prl2012}
\bibinfo{author}{\bibfnamefont{G.~J.} \bibnamefont{Baxter}},
  \bibinfo{author}{\bibfnamefont{S.~N.} \bibnamefont{Dorogovtsev}},
  \bibinfo{author}{\bibfnamefont{A.~V.} \bibnamefont{Goltsev}},
  \bibnamefont{and} \bibinfo{author}{\bibfnamefont{J.~F.~F.}
  \bibnamefont{Mendes}}, \bibinfo{journal}{Phys. Rev. Lett.}
  \textbf{\bibinfo{volume}{109}}, \bibinfo{pages}{248701}
  (\bibinfo{year}{2012}).

\bibitem[{\citenamefont{Danziger
  et~al.}(2014{\natexlab{b}})\citenamefont{Danziger, Bashan, Berezin,
  Shekhtman, and Havlin}}]{danziger-ndes2014}
\bibinfo{author}{\bibfnamefont{M.~M.} \bibnamefont{Danziger}},
  \bibinfo{author}{\bibfnamefont{A.}~\bibnamefont{Bashan}},
  \bibinfo{author}{\bibfnamefont{Y.}~\bibnamefont{Berezin}},
  \bibinfo{author}{\bibfnamefont{L.~M.} \bibnamefont{Shekhtman}},
  \bibnamefont{and} \bibinfo{author}{\bibfnamefont{S.}~\bibnamefont{Havlin}},
  in \emph{\bibinfo{booktitle}{Nonlinear Dynamics of Electronic Systems}},
  edited by \bibinfo{editor}{\bibfnamefont{V.}~\bibnamefont{Mladenov}}
  \bibnamefont{and} \bibinfo{editor}{\bibfnamefont{P.}~\bibnamefont{Ivanov}}
  (\bibinfo{publisher}{Springer International Publishing},
  \bibinfo{year}{2014}{\natexlab{b}}), vol. \bibinfo{volume}{438} of
  \emph{\bibinfo{series}{Communications in Computer and Information Science}},
  pp. \bibinfo{pages}{189--202}, ISBN \bibinfo{isbn}{978-3-319-08671-2}.

\bibitem[{\citenamefont{Centola}(2010)}]{centola-science2010}
\bibinfo{author}{\bibfnamefont{D.}~\bibnamefont{Centola}},
  \bibinfo{journal}{Science} \textbf{\bibinfo{volume}{329}},
  \bibinfo{pages}{1194} (\bibinfo{year}{2010}).

\bibitem[{\citenamefont{Weng et~al.}(2013)\citenamefont{Weng, Menczer, and
  Ahn}}]{weng-scireports2013}
\bibinfo{author}{\bibfnamefont{L.}~\bibnamefont{Weng}},
  \bibinfo{author}{\bibfnamefont{F.}~\bibnamefont{Menczer}}, \bibnamefont{and}
  \bibinfo{author}{\bibfnamefont{Y.-Y.} \bibnamefont{Ahn}},
  \bibinfo{journal}{Scientific Reports} \textbf{\bibinfo{volume}{3}},
  \bibinfo{pages}{2522} (\bibinfo{year}{2013}), ISSN \bibinfo{issn}{2045-2322}.

\bibitem[{\citenamefont{Nematzadeh et~al.}(2014)\citenamefont{Nematzadeh,
  Ferrara, Flammini, and Ahn}}]{nematzadeh-prl2014}
\bibinfo{author}{\bibfnamefont{A.}~\bibnamefont{Nematzadeh}},
  \bibinfo{author}{\bibfnamefont{E.}~\bibnamefont{Ferrara}},
  \bibinfo{author}{\bibfnamefont{A.}~\bibnamefont{Flammini}}, \bibnamefont{and}
  \bibinfo{author}{\bibfnamefont{Y.-Y.} \bibnamefont{Ahn}},
  \bibinfo{journal}{Phys. Rev. Lett.} \textbf{\bibinfo{volume}{113}},
  \bibinfo{pages}{088701} (\bibinfo{year}{2014}).

\bibitem[{\citenamefont{Cellai et~al.}(2013)\citenamefont{Cellai, L\'opez,
  Zhou, Gleeson, and Bianconi}}]{cellai-pre2013}
\bibinfo{author}{\bibfnamefont{D.}~\bibnamefont{Cellai}},
  \bibinfo{author}{\bibfnamefont{E.}~\bibnamefont{L\'opez}},
  \bibinfo{author}{\bibfnamefont{J.}~\bibnamefont{Zhou}},
  \bibinfo{author}{\bibfnamefont{J.~P.} \bibnamefont{Gleeson}},
  \bibnamefont{and} \bibinfo{author}{\bibfnamefont{G.}~\bibnamefont{Bianconi}},
  \bibinfo{journal}{Phys. Rev. E} \textbf{\bibinfo{volume}{88}},
  \bibinfo{pages}{052811} (\bibinfo{year}{2013}).

\bibitem[{\citenamefont{Li et~al.}(2013)\citenamefont{Li, Liu, Jia, and
  Wang}}]{li-newjphysics2013}
\bibinfo{author}{\bibfnamefont{M.}~\bibnamefont{Li}},
  \bibinfo{author}{\bibfnamefont{R.-R.} \bibnamefont{Liu}},
  \bibinfo{author}{\bibfnamefont{C.-X.} \bibnamefont{Jia}}, \bibnamefont{and}
  \bibinfo{author}{\bibfnamefont{B.-H.} \bibnamefont{Wang}},
  \bibinfo{journal}{New Journal of Physics} \textbf{\bibinfo{volume}{15}},
  \bibinfo{pages}{093013} (\bibinfo{year}{2013}).

\bibitem[{\citenamefont{Newman}(2010)}]{newman-book2010}
\bibinfo{author}{\bibfnamefont{M.}~\bibnamefont{Newman}},
  \emph{\bibinfo{title}{{Networks: an introduction}}} (\bibinfo{publisher}{OUP
  Oxford}, \bibinfo{year}{2010}).

\bibitem[{\citenamefont{Cohen and Havlin}(2010)}]{cohen-book2010}
\bibinfo{author}{\bibfnamefont{R.}~\bibnamefont{Cohen}} \bibnamefont{and}
  \bibinfo{author}{\bibfnamefont{S.}~\bibnamefont{Havlin}},
  \emph{\bibinfo{title}{Complex Networks: Structure, Robustness and Function}}
  (\bibinfo{publisher}{Cambridge University Press}, \bibinfo{year}{2010}), ISBN
  \bibinfo{isbn}{9781139489270}.

\bibitem[{\citenamefont{{Huang, Xuqing} et~al.}(2013)\citenamefont{{Huang,
  Xuqing}, {Shao, Shuai}, {Wang, Huijuan}, {Buldyrev, Sergey V.}, {Eugene
  Stanley, H.}, and {Havlin, Shlomo}}}]{huang-epl2013}
\bibinfo{author}{\bibnamefont{{Huang, Xuqing}}},
  \bibinfo{author}{\bibnamefont{{Shao, Shuai}}},
  \bibinfo{author}{\bibnamefont{{Wang, Huijuan}}},
  \bibinfo{author}{\bibnamefont{{Buldyrev, Sergey V.}}},
  \bibinfo{author}{\bibnamefont{{Eugene Stanley, H.}}}, \bibnamefont{and}
  \bibinfo{author}{\bibnamefont{{Havlin, Shlomo}}}, \bibinfo{journal}{EPL}
  \textbf{\bibinfo{volume}{101}}, \bibinfo{pages}{18002}
  (\bibinfo{year}{2013}).

\bibitem[{\citenamefont{Shao et~al.}(2014)\citenamefont{Shao, Huang, Stanley,
  and Havlin}}]{shao-pre2014}
\bibinfo{author}{\bibfnamefont{S.}~\bibnamefont{Shao}},
  \bibinfo{author}{\bibfnamefont{X.}~\bibnamefont{Huang}},
  \bibinfo{author}{\bibfnamefont{H.~E.} \bibnamefont{Stanley}},
  \bibnamefont{and} \bibinfo{author}{\bibfnamefont{S.}~\bibnamefont{Havlin}},
  \bibinfo{journal}{Phys. Rev. E} \textbf{\bibinfo{volume}{89}},
  \bibinfo{pages}{032812} (\bibinfo{year}{2014}).

\bibitem[{\citenamefont{Feng et~al.}(2008)\citenamefont{Feng, Deng, and
  Bl\"ote}}]{feng-pre2008}
\bibinfo{author}{\bibfnamefont{X.}~\bibnamefont{Feng}},
  \bibinfo{author}{\bibfnamefont{Y.}~\bibnamefont{Deng}}, \bibnamefont{and}
  \bibinfo{author}{\bibfnamefont{H.~W.~J.} \bibnamefont{Bl\"ote}},
  \bibinfo{journal}{Phys. Rev. E} \textbf{\bibinfo{volume}{78}},
  \bibinfo{pages}{031136} (\bibinfo{year}{2008}).

\end{thebibliography}
\end{document}